\documentclass[twocolumn,english,aps,prb,superscriptaddress,bibnotes,amsmath,amssymb,floatfix]{revtex4-1}
\usepackage[colorlinks=true,citecolor=blue,linkcolor=magenta]{hyperref}
\usepackage[utf8]{inputenc}
\usepackage[english]{babel}
\usepackage[T1]{fontenc}
\usepackage{textcomp}
\usepackage[markup=nocolor, authormarkupposition=left]{changes} 
\usepackage{soul}
\usepackage{url}
\usepackage{makecell}
\setaddedmarkup{{\color{red}#1}}
\usepackage{graphicx}
\usepackage{epstopdf}
\graphicspath{{./figures/}}
\makeatletter
 
\newcommand{\Rmnum}[1]{\expandafter\@slowromancap\romannumeral #1@} 
\makeatother
\usepackage{multirow}

\pdfcatalog{/OpenAction [1 0 R /FitH null]}

\begin{document}
\title{A scalable near-visible integrated photon-pair source for satellite quantum science}

\author{Yi-Han Luo}
\thanks{These authors contributed equally to this work.}
\affiliation{International Quantum Academy, Shenzhen 518048, China}

\author{Yuan Chen}
\thanks{These authors contributed equally to this work.}
\affiliation{International Quantum Academy, Shenzhen 518048, China}

\author{Ruiyang Chen}
\thanks{These authors contributed equally to this work.}
\affiliation{International Quantum Academy, Shenzhen 518048, China}
\affiliation{Hefei National Research Center for Physical Sciences at the Microscale and School of Physical Sciences, University of Science and Technology of China, Hefei 230026, China}

\author{Zeying Zhong}
\affiliation{International Quantum Academy, Shenzhen 518048, China}
\affiliation{Southern University of Science and Technology, Shenzhen 518055, China}

\author{Sicheng Zeng}
\affiliation{International Quantum Academy, Shenzhen 518048, China}
\affiliation{Southern University of Science and Technology, Shenzhen 518055, China}

\author{Baoqi Shi}
\affiliation{International Quantum Academy, Shenzhen 518048, China}
\affiliation{Hefei National Research Center for Physical Sciences at the Microscale and School of Physical Sciences, University of Science and Technology of China, Hefei 230026, China}

\author{Sanli Huang}
\affiliation{International Quantum Academy, Shenzhen 518048, China}
\affiliation{Hefei National Research Center for Physical Sciences at the Microscale and School of Physical Sciences, University of Science and Technology of China, Hefei 230026, China}

\author{Chen Shen}
\affiliation{International Quantum Academy, Shenzhen 518048, China}
\affiliation{Qaleido Photonics, Shenzhen 518048, China}

\author{Hui-Nan Wu}
\affiliation{Hefei National Research Center for Physical Sciences at the Microscale and School of Physical Sciences, University of Science and Technology of China, Hefei 230026, China}
\affiliation{Hefei National Laboratory, University of Science and Technology of China, Hefei 230088, China}
\affiliation{Shanghai Research Center for Quantum Sciences and CAS Center for Excellence in Quantum Information and Quantum Physics, University of Science and Technology of China, Shanghai 201315, China} 

\author{Yuan Cao}
\affiliation{Hefei National Research Center for Physical Sciences at the Microscale and School of Physical Sciences, University of Science and Technology of China, Hefei 230026, China}
\affiliation{Hefei National Laboratory, University of Science and Technology of China, Hefei 230088, China}
\affiliation{Shanghai Research Center for Quantum Sciences and CAS Center for Excellence in Quantum Information and Quantum Physics, University of Science and Technology of China, Shanghai 201315, China} 

\author{Junqiu Liu}
\email{liujq@iqasz.cn}
\affiliation{International Quantum Academy, Shenzhen 518048, China}
\affiliation{Hefei National Laboratory, University of Science and Technology of China, Hefei 230088, China}

\maketitle

\noindent \textbf{Quantum state distribution over vast distances is essential for global-scale quantum networks and fundamental test of quantum physics at space scale. 
While satellite platforms have demonstrated thousand-kilometer entanglement distribution, quantum key distribution and quantum teleportation with ground, future constellations and deep-space missions demand photon sources that are robust, compact, and power-efficient. 
Integrated photonics offers a scalable solution, yet a critical spectral gap persists.
Although telecom-band integrated photon-pair sources are well established, 
near-visible photons offer distinct advantages for satellite-to-ground links by mitigating diffraction loss and maximizing the collection efficiency of optical telescopes.
Scalable integrated sources in this regime have remained elusive due to the fundamental challenge of achieving anomalous dispersion in materials transparent at visible wavelengths.  
Here we bridge this gap by demonstrating an integrated near-visible photon-pair source based on a wide-bandgap, ultralow-loss, silicon nitride (Si$_3$N$_4$) microresonator. 
By engineering the dispersion of higher-order waveguide modes, we overcome the intrinsic normal dispersion limit to achieve efficient phase matching. 
The device exhibits a spectral brightness of 4.87$\times$10$^7$ pairs/s/mW$^2$/GHz and a narrow photon linewidth of 357 MHz.
We report high-purity heralded single-photon generation with a heralding rate up to 2.3 MHz and a second-order correlation function $g^{(2)}_\mathrm{h}(0)$ as low as 0.0041. 
Furthermore, we observe energy-time entanglement with 98.4\% interference visibility, violating the CHSH limit even at flux exceeding 40.6 million pairs/s. 
Combined with the proven radiation hardness of Si$_3$N$_4$, this source constitutes a flight-ready hardware foundation for daylight quantum communications and protocols requiring on-orbit multiphoton interference.
}

The distribution of quantum states over vast distances is fundamental to realizing global-scale quantum networks \cite{Kimble:08, Sidhu:21, Lu:2022, Gisin:07} and testing the interface between quantum mechanics and general relativity \cite{Vallone:16, Xuping:2019}. 
Photons are essential carriers for such missions, owing to their robustness against environmental decoherence. 
While satellite platforms have successfully demonstrated entanglement distribution and quantum teleportation over thousands of kilometers \cite{Yin:17, Chen:21, Ren:17, Liao:17}, these missions have relied on bulk optical setups that trade scalability for stability. 
To enable future satellite constellations \cite{Li:25a, Villar:20} and deep-space deployment, quantum light sources---specifically photon-pair sources that serve as the basis for heralded single photons and entanglement---must transition from bulky table-top assemblies to integrated solutions that are robust, compact, and power-efficient.

\begin{figure*}[t!]
\centering
\includegraphics[width=\linewidth]{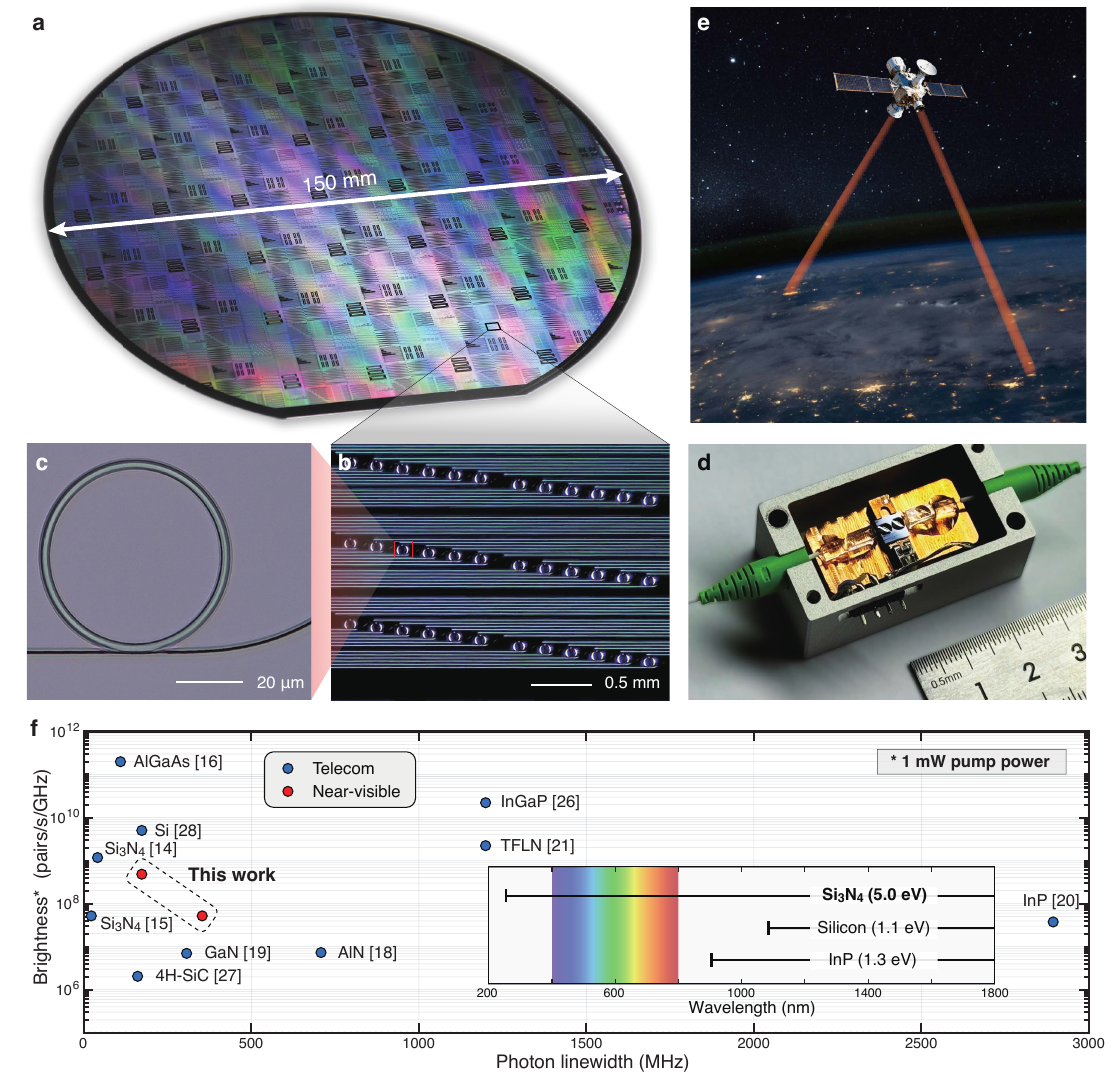}
\caption{
\textbf{Scalable manufacturing and satellite deployment of near-visible integrated photon-pair sources.}
\textbf{a}, 
Photograph of a 150-mm-diameter Si$_3$N$_4$ wafer fabricated using CMOS-compatible foundry processes, yielding hundreds of photonic chips.
\textbf{b}, 
Optical micrograph of a single chip hosting an array of microresonators, demonstrating high-density integration. 
\textbf{c}, 
Magnified view of the specific microresonator utilized for photon-pair generation in this work. 
\textbf{d}, 
The compact packaged module occupies a volume of only a few cubic centimeters, meeting the strict SWaP requirements for space payloads. 
\textbf{e}, 
Artistic rendering of the packaged source deployed on a quantum satellite. 
\textbf{f}, 
Performance benchmarking of state-of-the-art integrated photon-pair sources \cite{Chen:2024, Li:25, Steiner:21, Kumar:19, Guo:17a, Zeng:24, Zhao:22, Ma:20, Rahmouni:24, Yasui:25}. 
Spectral brightness (calibrated at 1 mW pump power) is plotted against photon linewidth. 
Blue and red circles denote sources operating in the telecommunication and near-visible bands, respectively.
Inset: Comparison of the bandgaps and transparency windows of common integrated photonic materials.  
The wide bandgap of Si$_3$N$_4$ (5.0 eV) makes it a unique candidate for high-performance operation in the near-visible band. 
}
\label{Fig:1}
\end{figure*}

Photonic integrated circuits (PICs) offer a pathway to meet these strict size, weight, and power (SWaP) requirements through scalable, CMOS-compatible manufacturing. 
As illustrated in Fig. \ref{Fig:1}, this compatibility enables high-density fabrication on 150-mm-diameter wafers, yielding hundreds of chips (Fig. \ref{Fig:1}a) that each host tens of robust microresonator sources (Fig. \ref{Fig:1}b, c).
These chips can be packaged into compact modules (Fig. \ref{Fig:1}d) occupying only a few cubic centimeters and weighing several hundred grams---an ideal form factor for satellite deployment (Fig. \ref{Fig:1}e).
Figure \ref{Fig:1}f summarizes the landscape of current integrated photon-pair sources realized across various material platforms \cite{Ma:17, Chen:2024, Li:25, Steiner:21, Pang:25, Guo:17a, Zeng:24, Kumar:19, Ma:20}, benchmarking their brightness and linewidths---the most critical performance metrics for quantum information processing. 
Comprehensive comparison and analysis are provided in Methods.
Within this landscape, silicon and Si$_3$N$_4$ have emerged as the dominant platforms, driven by their mature fabrication ecosystems \cite{Aghaee:25, Larsen:25, Alexander:25, Jia:25}.

However, a critical spectral mismatch exists between established integrated photonics and the requirements of satellite deployment. 
While integrated photon-pair sources are mature in the telecommunication bands, satellite links prioritize near-visible photons (typically 700--850 nm) to minimize beam divergence and maximize the collection efficiency of optical telescopes through diffraction-limited optics \cite{Lu:2022}. 
Furthermore, near-visible photons allow for efficient room-temperature detection using mature silicon detectors, enabling high-efficiency triggering for heralded single-photon downlinks. 

Achieving high-performance integrated photon-pair sources in this near-visible regime remains a formidable challenge. 
As illustrated in Fig. \ref{Fig:1}f inset, common materials such as silicon and indium phosphide (InP) feature narrow bandgaps and are opaque in the near-visible band. 
Conversely, while Si$_3$N$_4$ features a wide bandgap of 5.0 eV and is transparent, the material exhibits strong normal group velocity dispersion (GVD) in the near-visible regime, preventing the phase matching required for cavity-enhanced spontaneous four-wave mixing (SFWM). 
Here, we bridge this gap by demonstrating a flight-ready photon-pair source based on ultralow-loss Si$_3$N$_4$ microresonators. 
By engineering the GVD of higher-order waveguide modes, we overcome the normal GVD limit of the fundamental mode. 
Our device operates in the near-visible band and exhibits high spectral brightness and narrow photon linewidth, enabling the generation of heralded single photons and energy-time entangled states that meet the stringent requirements for satellite deployment. 

\begin{figure*}[t!]
\centering
\includegraphics[width=\linewidth]{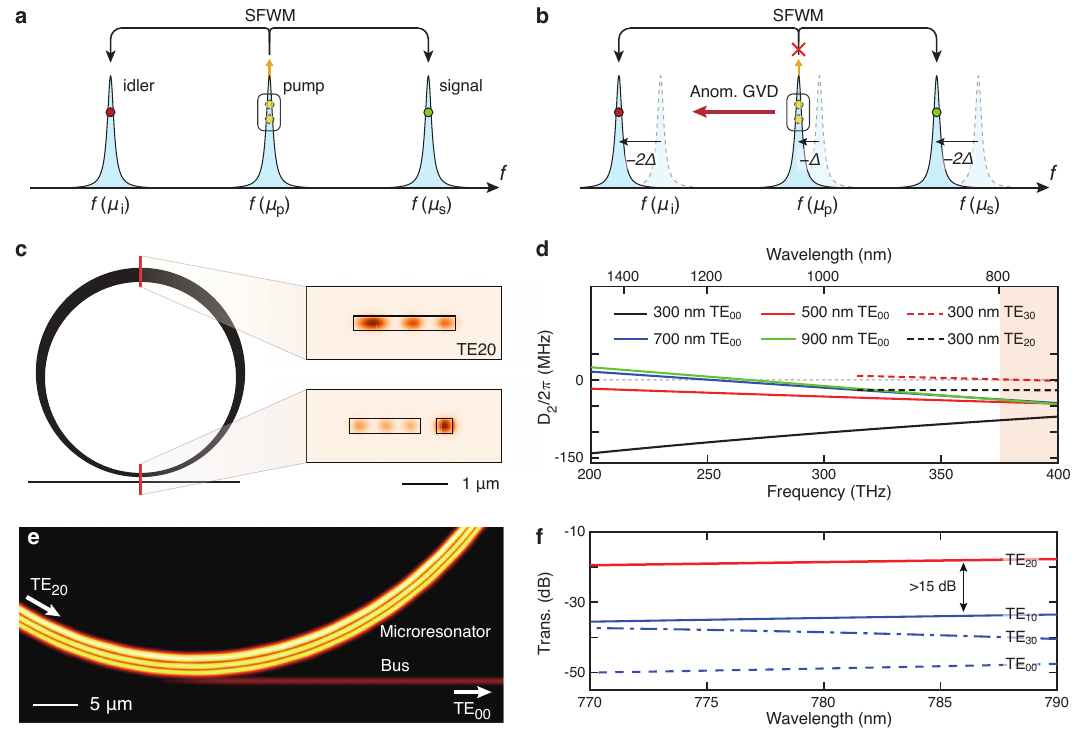}
\caption{
\textbf{Design and dispersion engineering of the near-visible photon-pair source.}
\textbf{a}, 
Principle of cavity-enhanced SFWM. 
Two pump photons at $f(\mu_\mathrm{p})$ annihilate to generate a signal–idler pair at $f(\mu_\mathrm{s})$ and $f(\mu_\mathrm{i})$, strictly aligned with the microresonator resonance grid. 
\textbf{b}, 
Diagram illustrating the phase-matching mechanism. 
Nonlinear SPM and XPM induce frequency detunings ($-\Delta$ and $-2\Delta$ for the pump and signal/idler resonances) that distort the resonance grid; 
efficient SFWM requires compensating for these shifts via anomalous GVD. 
\textbf{c}, 
Schematic of the Si$_3$N$_4$ microresonator featuring an adiabatically tapered waveguide geometry, accompanied by simulated optical modes. 
The waveguide width is expanded to 2200 nm in the loop to minimize propagation loss, and tapered to 1450 nm in the coupling region to ensure TE$_{20}$ excitation with near-unity coupling ideality. 
\textbf{d}, 
Simulated $D_2/2\pi$ for eigenmodes of a Si$_3$N$_4$ microresonator with 41.6 $\mu$m radius, 2200 nm width and varying thickness from 300 to 900 nm.
While the TE$_{00}$ mode exhibits strong normal GVD ($D_2<0$, solid lines) in the near-visible regime (orange shaded zoom) regardless of thickness, the higher-order modes TE$_{20}$ and TE$_{30}$ (dashed lines) achieve near-zero GVD. 
\textbf{e}, 
FDTD simulation of the coupling region, confirming near-unity coupling ideality from the microresonator TE$_{20}$ mode to the bus TE$_{00}$ mode.
The field magnitude is plotted as $|E|^{0.5}$ to enhance visualization. 
\textbf{f}, 
Simulated transmissions illustrating mode selectivity.  
The adiabatic coupler design suppresses parasitic excitation of modes other than the TE$_{20}$ by $>15$ dB.
}
\label{Fig:2}
\end{figure*}

\vspace{0.2cm}
\noindent \textbf{Device design}.
To satisfy the stringent link budgets of satellite-to-ground quantum channels, a photon-pair source must simultaneously exhibit high spectral brightness to overcome the severe channel attenuation (typically $>40$ dB), and narrow photon linewidth to facilitate the rejection of solar background noise. 
We employ SFWM in an integrated Si$_3$N$_4$ microresonator \cite{Helt:10}. 
As illustrated in Fig. \ref{Fig:2}a, a continuous-wave (CW) pump laser at frequency $f(\mu_\mathrm{p})$ drives the microresonator, where two pump photons annihilate to create a signal-idler pair at frequencies $f(\mu_\mathrm{s})$ and $f(\mu_\mathrm{i})$, governed by energy conservation $2f(\mu_\mathrm{p}) = f(\mu_\mathrm{s}) + f(\mu_\mathrm{i})$. 
Efficient SFWM requires phase matching; 
however, the nonlinear interaction induces self- and cross-phase modulation (SPM and XPM) that detunes the resonance grid as illustrated in Fig. \ref{Fig:2}b. 
This nonlinear shift prohibits phase matching and must be compensated by anomalous GVD. 

Achieving anomalous GVD in the near-visible band presents a fundamental material challenge. 
The microresonator dispersion is defined as 
\begin{equation} 
D_\mathrm{int} \equiv \omega(\mu) - \omega_0 - D_1\mu=D_2\mu^2/2 + \cdots
\end{equation} 
where $\omega(\mu)/2\pi$ is the $\mu$-th resonance frequency relative to the reference resonance of frequency $\omega_0/2\pi$, 
$D_1/2\pi$ is the free spectral range (FSR), 
and $D_2>0$ corresponds to anomalous GVD.  
Higher-order dispersion terms $\sum_{n=3}^{\cdots}D_n\mu^n/n!$  are neglected.
Although Si$_3$N$_4$ is transparent, the proximity of its ultraviolet bandgap imposes strong intrinsic normal GVD in the near-visible band (see the refractive index model in Methods). 
Consequently, conventional dispersion engineering via geometric confinement \cite{Okawachi:14}---while effective for the fundamental transverse-electric (TE$_{00}$) mode in the infrared---fails in this regime. 
Simulations in Fig. \ref{Fig:2}d show that the TE$_{00}$ mode remains in the normal GVD regime ($D_2<0$) even for waveguide thickness reaching 900 nm. 

We overcome this limitation by exploiting the dispersion properties of higher-order waveguide modes \cite{Kim:17, Zhao:20a}. 
As shown in Fig. \ref{Fig:2}d, higher-order modes such as TE$_{20}$ and TE$_{30}$ in 300-nm-thick waveguides exhibit near-zero GVD. 
Furthermore, mode coupling within these multimode waveguides creates avoided mode crossings (AMXs) \cite{Herr:14a}, which induce local deviation in the dispersion profile. 
We leverage these AMX-induced regions of local anomalous GVD to satisfy the phase-matching condition for SFWM.
To balance the trade-off between optical loss and dispersion, we select the TE$_{20}$ mode for following experiments.

The design strategy is further guided by the scaling laws of the photon-pair generation rate (PGR), which scales as 
\begin{equation}
    a\propto n_2^2Q^3/L^2 
\label{eqn:PGR}
\end{equation}
where $n_2$ is the nonlinear refractive index,  
$L$ is the cavity length,
and $Q=\omega/\kappa$ is the loaded quality factor. 
Simultaneously, the photon linewidth $\Delta\nu=0.64(\kappa_0 + \kappa_{\mathrm{ex}})/2\pi$ is dictated by the total cavity decay rate \cite{Ou:99}, where $\kappa_0$ and $\kappa_{\mathrm{ex}}$ are the intrinsic loss and external coupling rates. 
Theoretical derivation is provided in Supplementary Materials Note 1.
Therefore, maximizing PGR while minimizing linewidth requires a design that minimizes $\kappa_0$ while maintaining sufficient $\kappa_{\mathrm{ex}}$ for efficient photon extraction from the microresonator. 

We implement these criteria using an adiabatic waveguide geometry depicted in Fig. \ref{Fig:2}c. 
The microresonator features a radius of 41.6 $\mu$m and a thickness of 300 nm. 
To minimize optical scattering loss (related to $\kappa_0$), the waveguide width is expanded to 2200 nm in the region distal to the coupling region. 
In the coupling region, the width is adiabatically tapered to 1450 nm to optimize $\kappa_\mathrm{ex}$ to the 330-nm-wide bus waveguide. 
Finite-difference time-domain (FDTD) simulation in Fig. \ref{Fig:2}e confirms near-unity coupling ideality \cite{Pfeiffer:17} from the microresonator TE$_{20}$ mode to the bus TE$_{00}$ mode. 
The transmission simulations in Fig. \ref{Fig:2}f demonstrate that parasitic excitation of other microresonator modes is suppressed by $>15$ dB.

\begin{figure}[t!]
\centering
\includegraphics[width=\linewidth]{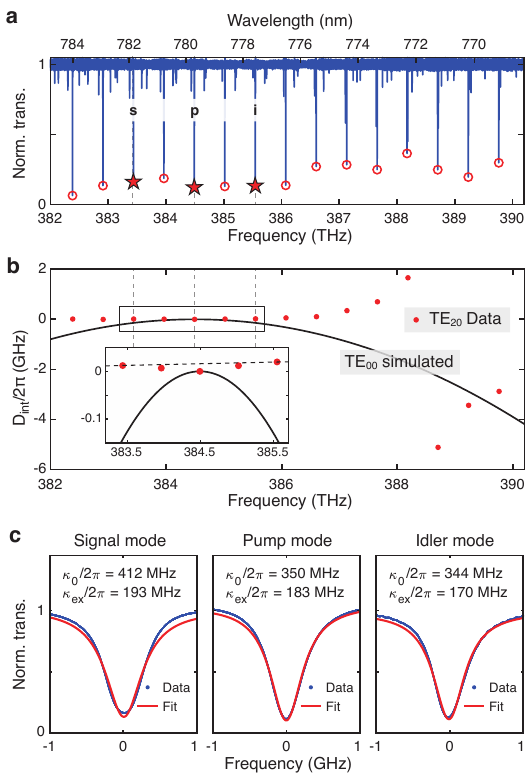}
\caption{
\textbf{Spectroscopic characterization and dispersion verification.}
\textbf{a}, 
Normalized microresonator transmission spectrum. 
Red stars mark the specific resonances selected for photon-pair generation: 
pump ($384.49233$ THz), signal ($383.43767$ THz), and idler ($385.54702$ THz).
\textbf{b}, 
Measured $D_\mathrm{int}/2\pi$ profile. 
The experimental data confirm the engineered anomalous GVD of the TE$_{20}$ mode, in sharp contrast to the normal GVD exhibited by the TE$_{00}$ mode (simulated, solid curve).
Inset: Magnified view of the phase-matching window. 
The pump mode frequency lies below the arithmetic mean of the signal and idler frequencies (dashed line), satisfying the phase-matching conditions required for SFWM. 
\textbf{c}, 
Resonance profiles of the signal, pump and idler modes, overlaid with their Lorentzian fits.
The extracted $\kappa_0/2\pi$ and $\kappa_\mathrm{ex}/2\pi$ are marked for each mode.
}
\label{Fig:3}
\end{figure}

\vspace{0.2cm}
\noindent \textbf{Device characterization}. 
We fabricate Si$_3$N$_4$ chip devices on 150-mm-diameter wafers in our CMOS foundry using a deep-ultraviolet subtractive process \cite{Ye:23, Sun:25}.
Stoichiometric Si$_3$N$_4$ films are deposited via low-pressure chemical vapor deposition (LPCVD) and subjected to high-temperature annealing to minimize optical loss and suppress fluorescence-induced background noise. 
Photographs of the wafer and the devices are shown in Fig. \ref{Fig:1}, and fabrication details are presented in Supplementary Materials Note 2. 
To verify the device physics, we characterize the microresonators using an atomic-transition-referenced vector spectrum analyzer \cite{Shi:25} with 161 kHz frequency resolution and 8.1 MHz frequency accuracy. 
Setup details are presented in Supplementary Materials Note 3.
Figure \ref{Fig:3}a shows the microresonator transmission spectrum spanning 382--390 THz, confirming the effective suppression of parasitic modes. 
We identify the target pump, signal, and idler resonances---marked by red stars---at 384.49233, 383.43767 and 385.54702 THz, respectively. 
By mapping the resonance frequencies $\omega(\mu)/2\pi$, we derive $D_\text{int}/2\pi$ as shown in Fig. \ref{Fig:3}b. 
Despite the wide FSR ($D_1/2\pi\approx527.4$ GHz), the $D_\text{int}$ profile unambiguously confirms the achievement of the designed near-zero GVD regime and the presence of AMXs.
Crucially, Fig. \ref{Fig:3}b inset box highlights that the pump mode lies below the linear interpolation between the signal and idler modes (dashed line), confirming the local anomalous GVD required for phase matching. 
In contrast, the simulated $D_\text{int}/2\pi$ of the TE$_{00}$ mode (solid curve) indicates a strong normal GVD.

The resonance profiles of the interacting TE$_{20}$ modes are detailed in Fig. \ref{Fig:3}c. 
Lorentzian fitting yields $(\kappa_0+\kappa_\text{ex})/2\pi$ values of $605$, $533$ and $514$ MHz for the signal, pump, and idler modes, respectively.  
These values correspond to loaded $Q$s of $0.63\times10^6$, $0.72\times10^6$ and $0.75\times10^6$, balancing high intra-cavity enhancement with efficient extraction.

\begin{figure*}[t!]
\centering
\includegraphics[width=\linewidth]{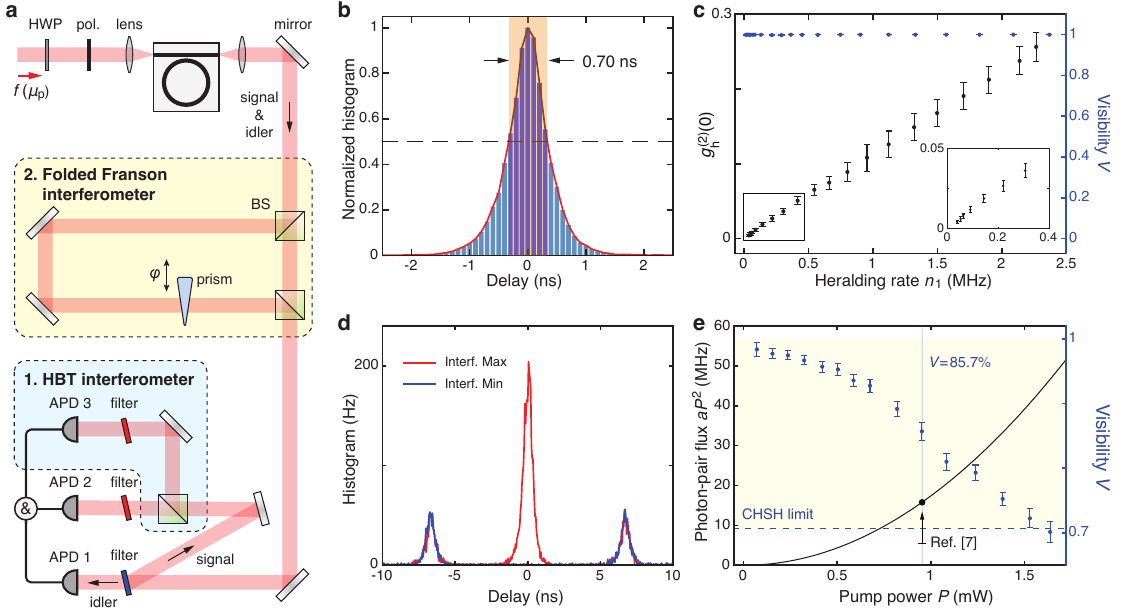}
\caption{
\textbf{Experimental configuration and performance characterization. } 
\textbf{a}, 
Schematic of the measurement setup. 
A CW pump laser of frequency $f(\mu_\mathrm{p})$ is coupled into the microresonator via aspherical lenses, and the generated photon pairs (signal and idler photons) are spectrally separated by narrow bandpass filters. 
The detection stage employs two interchangeable modules: 
an HBT interferometer for characterizing single-photon purity (Box 1) and a folded Franson interferometer for verifying energy-time entanglement (Box 2). 
HWP, half-wave plate; 
BS, beam splitter; 
APD, silicon avalanche photodiodes; 
pol., polarizer.
\textbf{b}, 
Normalized two-photon correlation histogram with 0.1 ns bin width showing a sharp temporal correlation peak with an FWHM of 0.70 ns. 
\textbf{c}, 
Heralded single-photon purity $g^{(2)}_\mathrm{h}(0)$ and visibility $V$ as a function of heralding rate $n_1$. 
The heralded $g^{(2)}_\mathrm{h}(0)$ remains low even as $n_1$ reaches 2.3 MHz at $P=1.5$ mW, confirming robust single-photon statistics. 
Inset: Magnified view of the zoom $g^{(2)}_\mathrm{h}(0)<0.05$.
\textbf{d}, 
Two-photon correlation histogram as a time-domain signature of energy-time entanglement. 
The central coincidence peak exhibits high contrast between constructive (red) and destructive (blue) two-photon interference. 
The bin width is 50 ps. 
\textbf{e}, 
Bell inequality violation and brightness benchmarking. 
The two-photon interference visibility $V$ consistently exceeds the classical CHSH limit of 0.707 (dashed line). 
At $P=1$ mW pump power, the device yields a photon-pair flux of $aP^2\approx15$ MHz while maintaining $V=85.7\%$, significantly outperforming benchmark bulk-optic sources ($\sim6$ MHz flux at $30$ mW pump in Ref. \cite{Yin:17}, marked by black arrow).
Error bars represent one standard deviation derived from Poissonian counting statistics. 
}
\label{Fig:4}
\end{figure*}

\noindent \textbf{Photon-pair generation}. 
We use the setup shown in Fig. \ref{Fig:4}a to generate and characterize photon pairs. 
Details are found in Methods and Supplementary Materials Note 4.
The resulting two-photon correlation histogram in Fig. \ref{Fig:4}b displays a sharp coincidence peak with a full-width at half-maximum (FWHM) of 0.70 ns, confirming photon-pair generation. 
By analyzing the scaling of coincidence and accidental count rates ($n_\mathrm{cc}$ and $n_\mathrm{acc}$) with on-chip pump power $P$, we extract a PGR of $a=1.74 \times 10^7$ pairs/s/mW$^2$ (see Methods). 
Combined with the narrow photon linewidth of $\Delta\nu=357$ MHz, this yields a spectral brightness of $a/\Delta\nu=4.87 \times 10^7$ pairs/s/mW$^2$/GHz. 
Calculation details are found in Methods and Supplementary Materials Note 1. 
Simultaneously, the source exhibits exceptional noise suppression; 
the coincidence-to-accidental ratio (CAR) reaches a maximum of $n_\mathrm{cc}/n_\mathrm{acc}-1=1061 \pm 21$ at $P=56$ $\mu$W (see Methods and Supplementary Materials Note 5).  

\vspace{0.2cm}
\noindent \textbf{Heralded single photons}. 
By detecting the idler photon to herald its signal counterpart, photon-pair sources function as practical single-photon emitters---essential resources for downlink quantum teleportation\cite{Bennett:93}, QKD \cite{Xu:20} and testing the quantum effects in curved spacetime \cite{Wu:24}. 
To characterize the single-photon purity of our source, we employ a Hanbury Brown and Twiss (HBT) interferometer marked in Fig. \ref{Fig:4}a Box 1. 
In this configuration, the registration of an idler photon at APD 1 heralds the presence of a signal photon, which is subsequently directed to a 50:50 beam splitter and monitored by APDs 2 and 3. 
We determine the heralded second-order correlation function $g^{(2)}_\mathrm{h}(0)=n_{123}n_1/(n_{12}n_{13})$, where $n_1$ represents the idler count rate at APD 1, and $n_{12}$, $n_{13}$, and $n_{123}$ are the coincidence rates for the APD combinations indicated by the subscripts. 

At the pump power of $P=153$ $\mu$W and the coincidence window width of 0.70 ns, we observe a pronounced photon anti-bunching with $g^{(2)}_\mathrm{h}(0)=0.0041 \pm 0.0010 $ at a heralding rate of $n_1=38$ kHz.
Crucially, as illustrated in Fig. \ref{Fig:4}c, the source maintains high purity even at elevated power; 
at $P=1.5$ mW, the heralding rate reaches $n_1=2.3$ MHz while $g^{(2)}_\mathrm{h}(0)$ remains below 0.3. 
This flux is sufficient to support deep-space quantum links; 
for instance, even accounting for the $\sim40$ dB channel loss \cite{Cao:18} expected for a downlink from a stable Lagrange point to Earth, our source still provides a ground-based detection rate exceeding 100 Hz. 
Beyond this regime, the system transitions from quantum SFWM to classical optical parametric oscillation (OPO), signified by an abrupt rise in $g^{(2)}_\mathrm{h}(0)$ (see Methods). 
Finally, as shown in Fig. \ref{Fig:4}c, self-interference measurements of the heralded signal photons yield near-unity visibility, confirming the high coherence required for testing the quantum effects in curved spacetime \cite{Wu:24}. 

\vspace{0.2cm}
\noindent \textbf{Energy-time entanglement}. 
The generated photon pairs exhibit intrinsic energy-time entanglement, a degree of freedom exceptionally robust against decoherence and ideal for long-distance quantum distribution \cite{Franson:89}. 
Under CW pumping, the coherent superposition of photon pairs generated at distinct times $|e_\mathrm{s}, e_\mathrm{i}\rangle$ (early) and $|l_\mathrm{s}, l_\mathrm{i}\rangle$ (late) yields an entangled state:
\begin{equation}
    |\Psi\rangle = \frac{1}{\sqrt{2}}\left( |e_\mathrm{s}, e_\mathrm{i}\rangle + |l_\mathrm{s}, l_\mathrm{i}\rangle \right)
\end{equation}
To verify this entanglement and resolve two-photon interference, we employ a folded Franson interferometer as depicted in Fig. \ref{Fig:4}a Box 2. 
The photons are equally split into two arms with a path length difference of 2 m, introducing a temporal delay of $\sim6.6$ ns. 
This delay significantly exceeds the two-photon correlation histogram FWHM (0.70 ns), effectively erasing single-photon interference.  
The signal and idler photons are projected onto the superposition state $|\phi\rangle = \left(|e\rangle + e^{i\varphi} |l\rangle\right)/\sqrt{2}$, where a small-angle wedge prism controls the relative phase $\varphi$. 
The two-photon interference---measuring the state $|\Psi\rangle$ in the state $|\phi_\mathrm{s}\rangle|\phi_\mathrm{i}\rangle$---is governed by the probability $p=(1+\cos 2\varphi)/4$. 

The resulting interference is captured in the correlation histogram presented in Fig. \ref{Fig:4}d. 
We observe high-contrast modulation of the central coincidence peak: 
The central peak is fourfold of the sidebands at $\varphi = 0$, and completely vanishes at $\varphi = \pi/2$ due to destructive interference.
To ensure efficient post-selection, we use a coincidence window of 3.0 ns width for the central peak to obtain $n_\mathrm{cc}$.
As $\varphi$ varies, $n_\mathrm{cc}$ oscillates sinusoidally with a period of $\pi$, with the maximum and minimum denoted as $n_\mathrm{cc}^\mathrm{max}$ and $n_\mathrm{cc}^\mathrm{min}$. 
The raw two-photon interference visibility---calculated as $V=(n_\mathrm{cc}^\mathrm{max}-n_\mathrm{cc}^\mathrm{min})/(n_\mathrm{cc}^\mathrm{max}+n_\mathrm{cc}^\mathrm{min})$ without background subtraction---reaches $98.4\% \pm 1.1\%$ at $P=71$ $\mu$W. 
Figure \ref{Fig:4}e illustrates the pump-power-dependent visibility $V$ and corresponding photon-pair flux (calculated as $aP^2$). 
Notably, $V$ remains consistently above the classical CHSH limit of 0.707 (dashed line) across the entire power range up to the OPO threshold. 
The interference fringes and the histograms for various pump powers are provided in Supplementary Materials Note 5.

This performance represents a significant leap over existing satellite-based systems. 
Compared to the bulk-optic source used in the landmark 1,200-km entanglement distribution \cite{Yin:17}, our integrated device delivers 2.5 times the photon-pair flux (15 MHz vs. 6 MHz) at comparable visibility ($V=85.7\%$) while consumes only $3\%$ of the pump power (1 mW vs. 30 mW).
Even accounting for the severe 70-dB channel loss characteristic of dual-downlink satellite transmission, this 15 MHz photon flux ensures a ground-based coincidence count rate exceeding 1 Hz, confirming the device's viability for future space-to-ground quantum networks.

\noindent \textbf{Discussion and outlook. }
Our integrated source bridges the critical spectral gap required for satellite quantum links, achieving brightness and linewidth metrics in the near-visible regime that rival mature telecom-band devices. 
As summarized in Fig. \ref{Fig:1}f and Extended Data Table \ref{tab:S1}, our platform delivers a superior combination of scalability and performance compared to existing spontaneous parametric down-conversion (SPDC) and SFWM sources, including those based on bulk free-space optics.

We further explore the performance limits of our platform using a microresonator where AMXs with the TM$_{10}$ mode induce favorable local anomalous GVD. 
This device exhibits an extreme spectral brightness of $5.84\times10^8$ pairs/s/mW$^2$/GHz and a photon linewidth of just 159 MHz. 
Details are found in Supplementary Materials Note 5. 
The lower OPO threshold inherent to it limits the maximum attainable photon-pair flux (theoretically $\propto Q^{-1}$ and independent with nonlinear refractive index $n_2$,  see Methods and Supplementary Materials Note 1). 
Nevertheless, such narrow linewidths are particularly attractive for interfacing with atomic quantum memories \cite{Pang:20}.
Furthermore, this linewidth enables highly efficient multi-photon interference using nanosecond pulses, offering a robust and space-compatible alternative to conventional broadband sources that rely on complex femtosecond lasers \cite{Ren:17}.

To transition from chip-scale demonstration to flight-ready payload, we define a clear path for system-level integration. 
Optimizing the inverse taper geometry is expected to boost fiber coupling efficiency beyond 70\% with standard single mode fibers (780-HP, see Supplementary Materials Note 6).   
In addition, by leveraging self-injection locking with near-visible laser chips \cite{Kondratiev:23, Long:25a}, the source can evolve into a hybrid or heterogeneous integrated module \cite{Sun:25, Xiang:21}.
This architecture would yield a turnkey device occupying only a few cubic centimeters while operating on an electrical power budget of less than 5 W---meeting the strict SWaP constraints of satellite constellations. 
A conceptual schematic of this integrated system is provided in Supplementary Materials Note 6.
Finally, the platform offers exceptional thermal stability; 
given the low thermo-optic coefficient of Si$_3$N$_4$ (resonance shift of $\sim4$ GHz/K in the near-visible), regulating the chip temperature to within 10 mK stabilizes the frequency to 40 MHz, significantly simplifying thermal management in the harsh environment of space.

In summary, we have demonstrated a CMOS-compatible, near-visible photon-pair source based on a high-$Q$ Si$_3$N$_4$ microresonator chip. 
The scalable manufacturing of these integrated devices promises to catalyze the rapid deployment of quantum microsatellite constellations \cite{Li:25a, Villar:20}. 
Simultaneously, our device combines high spectral brightness and narrow photon linewidth, enabling high-purity heralded single photons with $g_\mathrm{h}^{(2)}(0) < 0.005$ and high-visibility energy-time entanglement with $V > 0.98$.
By bridging such superior performance with a compact, scalable solid-state architecture, this work establishes a viable hardware foundation for next-generation global quantum networks and space-borne quantum technologies.

\medskip

\setcounter{table}{0} 
\renewcommand{\tablename}{Extended Data Table}
\renewcommand{\thetable}{\arabic{table}}

\setcounter{figure}{0} 
\renewcommand{\figurename}{Extended Data Figure}
\renewcommand{\thefigure}{\arabic{figure}}

\vspace{0.3cm}
\noindent \textbf{Methods}

\noindent\textbf{Refractive index model for simulations. }
To ensure accurate $D_\mathrm{int}$ simulations, the refractive indices of the LPCVD SiO$_2$ and Si$_3$N$_4$ are characterized via spectroscopic ellipsometry and fitted to a three-term Sellmeier model

\begin{subequations}
    \begin{align}
        \notag n_\mathrm{{Si_3N_4}}^2 &= 1 + \frac{1.73783 \lambda^2}{\lambda^2 - 0.00745}
        + \frac{1.25000\lambda^2}{\lambda^2 - 0.03128}\\
        &~~~~~~+ \frac{2.68390\lambda^2}{\lambda^2 - 143.75857},\\
        \notag n_\mathrm{{SiO_2}}^2 &= 1 + \frac{0.69652\lambda^2}{\lambda^2 - 0.00429}
        + \frac{0.40321\lambda^2}{\lambda^2 - 0.01401}\\
        &~~~~~~+ \frac{0.85807\lambda^2}{\lambda^2 - 96.57021}, 
    \end{align}
\end{subequations}
where the wavelength $\lambda$ is expressed in micrometers.

\begin{figure*}[t!]
\centering
\includegraphics[width=\linewidth]{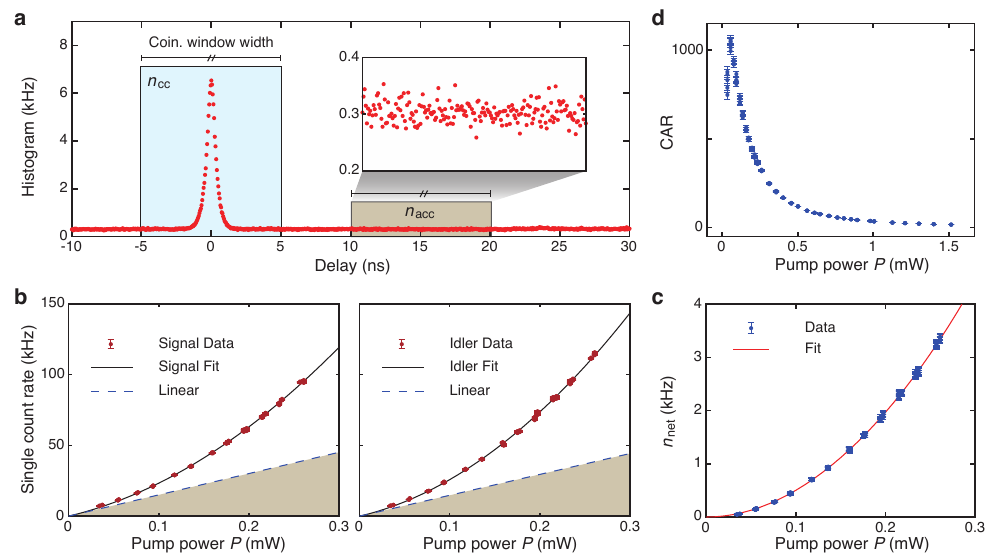}
\caption{
\textbf{Characterization of PGR and CAR.}
\textbf{a}, 
Experimental definition of the coincidence count rate $n_\mathrm{cc}$ and accidental coincidence count rate $n_\mathrm{acc}$. 
A representative two-photon correlation histogram (recorded at $P=1.52$ mW) is presented to illustrate the extraction procedure in the presence of background noise.
The $n_\mathrm{cc}$ is determined by integrating the histogram bins within a coincidence window centered at zero delay, while the $n_\mathrm{acc}$ is calculated using an identical window at a large delay offset.
The coincidence window width is adaptable, allowing for optimization across different experimental regimes.
\textbf{b}, 
Measured single-photon count rates for signal ($n_\mathrm{s}$) and idler ($n_\mathrm{i}$) modes as a function of pump power $P$. 
Data points are fitted with polynomial functions (solid curves). 
The linear components (dashed lines) represent the background noise floor.
\textbf{c}, 
Net coincidence count rate $n_\mathrm{net}$ as a function of $P$, exhibiting the expected quadratic scaling confirmed by a polynomial fit.
\textbf{d}, 
Measured CAR as a function of $P$.
Error bars represent one standard deviation derived from Poissonian counting statistics. 
}
\label{Fig:E1}
\end{figure*}

\vspace{0.2cm}
\noindent\textbf{Experimental setup.}
The microresonator is pumped via free-space coupling as illustrated in Fig. \ref{Fig:4}a. 
We align the pump polarization to the $\text{TE}_{00}$ mode of the bus waveguide using a half-wave plate (HWP) and a polarizer. 
Input and output coupling of the chip are facilitated by aspherical lenses with 2.8 mm focal length. 
The generated signal and idler photons are spatially separated using a series of bandpass filters ($0.7$-nm bandwidth) centered at their respective resonance frequencies. 
Photons are detected by silicon avalanche photodiodes (APDs, Excelitas SPCM-NIR) and processed by a time tagger (Swabian Instruments Time Tagger Ultra). 
Detailed component specifications are provided in Supplementary Materials Note 4.

\vspace{0.2cm}
\noindent\textbf{Measurement of PGR. }
Experimentally, the single-photon count rates $n_\mathrm{s/i}$ are detected by APDs and recorded directly by the time tagger. 
Specifically, $n_1=n_\mathrm{i}$ is the heralding rate in heralded single-photon generation. 
As illustrated in Extended Data Fig. \ref{Fig:E1}a, the coincidence count rate $n_\mathrm{cc}$ is determined by integrating histogram bins within a window centered at zero delay, while the accidental coincidence rate $n_\mathrm{acc}$ is calculated using an identical window width but at a large delay offset. 
The net coincidence rate is defined as $n_\mathrm{net} = n_\mathrm{cc} - n_\mathrm{acc}$.
The photon-pair flux $\alpha$ scales quadratically with pump power ($\alpha = aP^2$), whereas the noise-photon flux $\beta_\mathrm{s/i}$ scales linearly, i.e. $\beta_\mathrm{s/i} = b_\mathrm{s/i}P$ where $b_\mathrm{s/i}$ is the noise-photon generation rate. 
Consequently, the measured rates are modeled as:

\begin{align}
    n_\mathrm{s/i} &= \eta_\mathrm{s/i} (aP^2 + b_\mathrm{s/i}P), \\
    n_\mathrm{net} &= \eta_\mathrm{i}\eta_\mathrm{s} aP^2 + \mathcal{O}(P^3), 
\end{align}
where $\eta_\mathrm{s/i}$ represents the total system efficiency for the signal and idler channels. 
Utilizing these distinct power-scaling laws, PGR can be accurately extracted. 

Experimentally, we vary $P$ to record $n_\mathrm{s/i}$ and $n_\mathrm{net}$ values, as presented in Extended Data Fig. \ref{Fig:E1}b and Fig. \ref{Fig:E1}c. 
To ensure accurate brightness estimation and capture all coincidence events without overestimation \cite{Chen:2024}, we adopt a coincidence window of 20 ns width.
Through polynomial fitting, we extract the combined coefficients $\eta_{\mathrm{s}}a$, $\eta_{\mathrm{i}}a$, and $\eta_{\mathrm{s}}\eta_{\mathrm{i}}a$, allowing for simultaneous determination of the channel efficiencies $\eta_\mathrm{s/i}$ and PGR ($a$). 
The linear components of $n_\mathrm{s/i}$ (dashed lines in Extended Data Fig. \ref{Fig:E1}b) represent the background noise floor. 
Accounting for the measured overall system efficiencies ($\eta_\mathrm{s}=4.8\%$, $\eta_\mathrm{i}=6.4\%$), we obtain a PGR of $a=1.74 \times 10^7$ pairs/s/mW$^2$. 
A detailed system loss budget is provided in Supplementary Materials Note 5.

\begin{figure*}[t!]
\centering
\includegraphics[width=\linewidth]{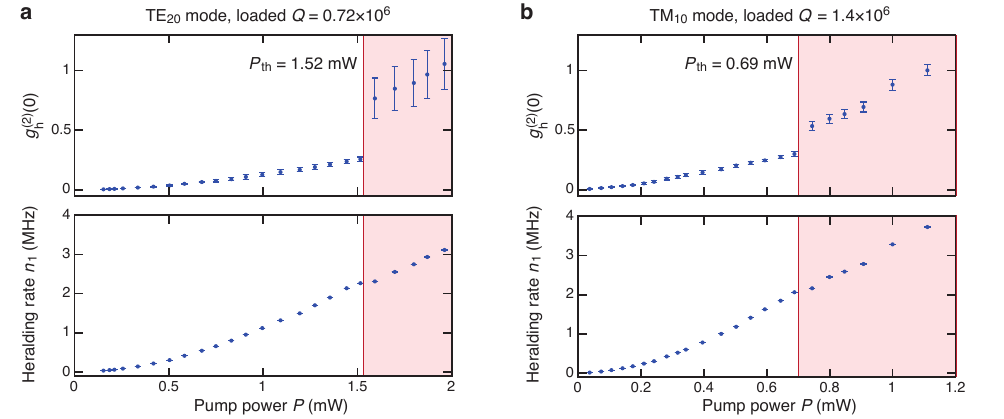}
\caption{
\textbf{Observation of the transition from SFWM to OPO}. 
As pump power $P$ increases, $g_\mathrm{h}^{(2)}(0)$ rises abruptly toward unity, signaling the transition from SFWM to OPO. 
The power where this transition occurs is defined as the oscillation threshold $P_\mathrm{th}$.
\textbf{a}, 
Measured $g_\mathrm{h}^{(2)}(0)$ and heralding rate $n_1$ for the TE$_{20}$ mode (loaded $Q = 0.72 \times 10^6$) as a function of $P$ up to 2 mW.
At $P_\mathrm{th} = 1.52$ mW, the heralding rate is $n_1=2.3$ MHz.
\textbf{b}, 
Measured $g_\mathrm{h}^{(2)}(0)$ and heralding rate $n_1$ for the TM$_{10}$ mode (loaded $Q = 1.4 \times 10^6$) as a function of $P$ up to 1.2 mW.
At $P_\mathrm{th} = 0.69$ mW, the heralding rate is $n_1=2.1$ MHz.
Error bars represent one standard deviation derived from Poissonian counting statistics. 
}
\label{Fig:E2}
\end{figure*}

\vspace{0.2cm}
\noindent\textbf{Photon linewidth calculation}.
Following the theoretical framework in Ref. \cite{Ou:99}, the photon linewidth for cavity-enhanced SFWM is $\Delta \nu=0.64\kappa/2\pi$ where $\kappa=\kappa_0+\kappa_\text{ex}$. 
For a microresonator where the signal and idler modes exhibit distinct decay rates ($\kappa_s$ and $\kappa_i$), the photon linewidth is calculated as
\begin{equation}
\Delta\nu = \frac{1}{2\pi\sqrt{2}}\sqrt{\sqrt{(\kappa_\mathrm{s}^2+\kappa_\mathrm{i}^2)^2 + 4\kappa_\mathrm{s}^2\kappa_\mathrm{i}^2}-(\kappa_\mathrm{s}^2+\kappa_\mathrm{i}^2)}
\end{equation}
Substituting the experimentally extracted decay rates $\kappa_\mathrm{s}/2\pi=605$ MHz and $\kappa_\mathrm{i}/2\pi=514$ MHz, we calculate $\Delta\nu=357$ MHz. 
Detailed theoretical derivation is provided in Supplementary Materials Note 1.

\vspace{0.2cm}
\noindent\textbf{Measurement of CAR}.
The rates $n_\mathrm{cc}$ and $n_\mathrm{acc}$ are extracted over a pump power range up to $P=1.5$ mW. 
Extended Data Fig. \ref{Fig:E1}d shows the measured CAR. 
The CAR reaches the maximum of $n_\mathrm{cc}/n_\mathrm{acc}-1=1061 \pm 21$ at $P=56$ $\mu$W.
The observed power dependence follows the expected behavior for SFWM sources: 
at low power the CAR is limited by the detector's dark counts ($\sim 500$ Hz), while at higher power the CAR degrades by the increasing probability of multi-photon emission. 
Detailed noise analysis is provided in Supplementary Materials Note 5.

\vspace{0.2cm}
\noindent\textbf{OPO threshold and flux limit. }
As the pump power $P$ increases, we observe an abrupt change of $g_\mathrm{h}^{(2)}(0)$ toward unity, as shown in Extended Data Fig. \ref{Fig:E2}a. 
This behavior signifies a transition from a non-classical single-photon state to a Poissonian distribution, indicating the onset of coherent light emission. 
We define $P$ at this transition as the OPO threshold power $P_\mathrm{th}$. 
When $P>P_\mathrm{th}$, the generated photons undergo parametric amplification within the microresonator, marking the regime where the photons behave classically. 
Given that $P_\mathrm{th}\propto Q^{-2}$ (Ref. \cite{Pu:16}), and PGR $\propto Q^{3}$ (see Eq. \ref{eqn:PGR}), the theoretical maximum achievable photon-pair flux $aP^2_\mathrm{th}\propto Q^{-1}$. 
This suggests a fundamental trade-off: a higher $Q$-factor does not inherently enable a higher maximum photon-pair flux.
This behavior is experimentally corroborated by comparing with results of $\text{TM}_{10}$ mode, which possesses a higher loaded $Q=1.4 \times 10^6$.
As shown in Extended Data Fig. \ref{Fig:E2}b, the transition occurs at a lower $P_\mathrm{th} = 0.69$ mW, yielding a maximum heralding rate of $2.1$ MHz. 
A comprehensive analysis incorporating the specific contributions of $\kappa_0$ and $\kappa_\mathrm{ex}$ is provided in Supplementary Materials Note 1.

\begin{figure}[t!]
\centering
\includegraphics[width=\linewidth]{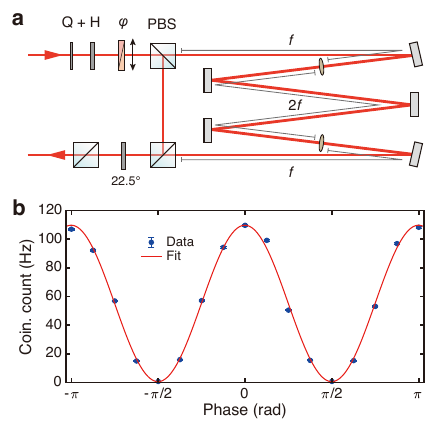}
\caption{
\textbf{Folded Franson interferometer and the two-photon interference fringe}.
\textbf{a}, 
Schematic of the free-space folded Franson interferometer implementation. 
\textbf{b}, 
Measured two-photon interference fringe fitted with a sinusoidal function (red curve), yielding a visibility of $V=98.7\%$. 
Error bars represent one standard deviation derived from Poissonian counting statistics, which are however much smaller than the data point size. 
}
\label{Fig:E3}
\end{figure}

\vspace{0.2cm}
\noindent\textbf{Folded Franson interferometer}. 
A folded Franson interferometer---an unbalanced Mach-Zehnder interferometer---is employed to resolve the two-photon interference visibility of the energy-time entanglement. 
As shown in Extended Data Fig. \ref{Fig:E3}a, the polarization of incident photons is precisely adjusted using a combination of a quarter-wave plate (QWP) and an HWP---denoted as ``Q+P''---to ensure the flux is equally split by the subsequent polarizing beam splitter (PBS).
Following the first PBS, the photons are distributed into two arms with a path length difference of 2 m, corresponding to a temporal delay of approximately 6.6 ns. 
This delay is designed to significantly exceed the two-photon correlation histogram FWHM (0.7 ns), thereby effectively erasing the single-photon interference. 
The two paths are recombined at a second PBS and projected onto the superposition state $(|H\rangle+|V\rangle)/\sqrt{2}$ to observe the two-photon interference.
To control the relative phase $\varphi$, a phase shifter is utilized before the first PBS. 
This phase shifter consists of two angled quartz plates glued together, whose optical axes are oriented perpendicular to each other. 
To correct for spatial mode mismatch caused by the vastly different propagation lengths at the second PBS, a 4-$f$ optical system consisting of two lenses with a 400-mm focal length is integrated into the long arm. 
This system ensures that the spatial mode transformation corresponding to the long arm approximates the identity matrix, thereby maintaining high interference visibility.

\vspace{0.2cm}
\noindent\textbf{Two-photon interference fringe}.
The two-photon interference fringe characterizing the energy-time entanglement is presented in Extended Data Fig. \ref{Fig:E3}b. 
The measurement is performed at a pump power of $P=164$ $\mu$W with a coincidence window width of 3.0 ns to ensure the capture of all coincidence events. 
Each data point represents the coincidence count rate $n_\mathrm{cc}$ averaged over 30 seconds without background subtraction. 
The observed fringe oscillates with a period of $\pi$, a distinct signature of two-photon interference. 
We extract visibility by fitting the data to the sinusoidal model $n_\mathrm{cc} = 0.5A[1 + V \cos 2\varphi]$, yielding a raw visibility of $V = (98.7\pm1.4)\%$. 
This result is consistent with the visibility calculated directly from the interference maxima and minima.

\vspace{0.2cm}
\noindent\textbf{Comparisons with other photon-pair sources.}
Comprehensive performance metrics for photon-pair sources across various material platforms are summarized in Extended Data Table \ref{tab:S1}, alongside their respective CMOS compatibility. 
For the present work, we report performance metrics specifically for the TE$_{20}$ and TM$_{10}$ modes.
In cases where the literature reports microresonator-based generation rates but lacks explicit brightness values, we estimate the spectral brightness using the photon linewidth derived as $0.64\kappa/2\pi$. 
Furthermore, to provide a broader context for the state of the art, we also benchmark our results against two landmark studies employing free-space optics for satellite-based (810 nm wavelength) and fiber-based (1560 nm wavelength) quantum networks.

\medskip
\begin{footnotesize}

\noindent \textbf{Acknowledgments}: 
We thank Jian-Wei Pan, Cheng-Zhi Peng and Dapeng Yu for advices and support, 
Jinbao Long, Ying Zuo and J. F. Chen for helping the experiment, 
Zhenyuan Shang and Jiahao Sun for helping the chip fabrication, 
Yue Hu for helping the chip characterization, 
and Zheng-Da Li for fruitful discussion. 
We acknowledge support from the National Key R\&D Program of China (Grant No. 2024YFA1409300), 
National Natural Science Foundation of China (Grant No. 12404417 and U25D9005), 
Quantum Science and Technology--National Science and Technology Major Project (Grant No. 2023ZD0301500), 
Shenzhen Science and Technology Program (Grant No. RCJC20231211090042078), 
Guangdong-Hong Kong Technology Cooperation Funding Scheme (Grant No. 2024A0505040008), 
and Shenzhen-Hong Kong Cooperation Zone for Technology and Innovation (HZQB-KCZYB2020050). 
Y.-H L. acknowledges support from the Young Elite Scientists Sponsorship Program by CAST (Grant No. YESS20240475). 
The silicon nitride chips were fabricated by Qaleido Photonics. 

\noindent \textbf{Author contributions}: 
Y.-H. L. and J. L. conceived the experiment. 
Y.-H. L. designed the chip devices. 
Z. Z., S. H. and C. S. fabricated the chip devices. 
Y. Chen, S. Z. and B. S. characterized the chip devices. 
Y. Chen, R. C. and Y.-H. L. built the experimental setup and performed the experiments, assisted by H.-N. W. and Y. Cao. 
Y.-H. L., Y. Chen, R. C. and J. L. analyzed the data and prepared the manuscript with input from others. 
J. L. supervised the project.  

\noindent \textbf{Conflict of interest}:
C. S. and J. Liu are co-founders of Qaleido Photonics, a start-up that is developing heterogeneous silicon nitride integrated photonics technologies.  Others declare no conflicts of interest.

\noindent \textbf{Data Availability Statement}: 
The code and data used to produce the plots within this work will be released on the repository \texttt{Zenodo} upon publication of this preprint.

\end{footnotesize}

\onecolumngrid

\begin{table*}[h!]
\caption{
\textbf{Performance benchmark of integrated photon-pair sources.}
Comprehensive comparison of performance metrics for photon-pair sources. 
Performance metrics for this study are specified for the TE$_{20}$ and TM$_{10}$ modes. 
TPI, two-photon interference.
}
\begin{ruledtabular}
\begin{tabular}{l@{}ccccccccc}
\multirow{2}*{Reference} & \multirow{2}*{Material} & Wavelength & CMOS & \multirow{2}*{Type} &PGR@1 mW & \multirow{2}*{Loaded $Q$} & Linewidth & Brightness@1 mW & TPI \\
 & & (nm) & compatible &  & (Hz) &  & (GHz) & (Hz/GHz) &  visibility\\[1.5pt]
\colrule\\[-8pt]
This work \footnote{Using TE$_{20}$ mode} & Si$_3$N$_4$ & 780 & Yes & SFWM & $1.74\times10^7$ & $0.7\times10^6$ & 0.36 & $4.87\times10^7$ & 98.4\% \\[1.6pt]
This work \footnote{Using TM$_{10}$ mode} & Si$_3$N$_4$ & 780 & Yes & SFWM & $9.29\times10^7$ & $1.4\times10^6$ & 0.16 & $5.84\times10^8$ & 97.9\% \\[1.6pt]
\colrule\\[-6pt]
Chen \emph{et al}. \cite{Chen:2024} & Si$_3$N$_4$ & 1550 & Yes & SFWM & $3.04\times10^7$ & $5.0\times10^6$ & 0.026 & $1.2\times10^9$ & 97.3\% \\[1.6pt]
Li \emph{et al}. \cite{Li:25} & Si$_3$N$_4$ & 1560 & Yes & SPDC & $5.3\times10^5$ & $1.4\times10^7$ & 0.009 & $6.0\times10^7$ & -- \\[1.6pt]
Fan \emph{et al}. \cite{Fan:23} & Si$_3$N$_4$ & 1550 & Yes & SFWM & $1.15\times10^6$ & $1.0\times10^6$ & 0.12 & $9.6\times10^6$ & 97.3\% \\[1.5pt]%
Lu \emph{et al}. \cite{Lu:19a} & Si$_3$N$_4$  & 1548+668 & Yes & SFWM & $2.3\times10^6$ & $1.0\times10^6$ & 0.91 & $1.1\times10^7$ & 82.7\% \\[1.5pt]
Ma \emph{et al}. \cite{Ma:17} & Silicon & 1550 & Yes & SFWM & $1.5\times10^8$ & $9.2\times10^4$ & 2.1 & $1.6\times10^8$ & 95.9\% \\[1.5pt]
Yasui \emph{et al}. \cite{Yasui:25} & Silicon & 1536 & Yes & SFWM & $1.29\times10^9$ & $6.5\times10^5$ & 0.19 & $6.77\times10^9$ & 98.0\% \\[1.5pt]
Rahmouni \emph{et al}. \cite{Rahmouni:24} & 4H-SiC & 1550 & Yes & SFWM & $3.1\times10^5$ & $8\times10^5$ & 0.15 & $2.1\times10^6$ & 99.2\% \\[1.5pt]
Li \emph{et al.} \cite{LiJ:24} & 3C-SiC & 1550 & Yes & SPDC & 9.6$\times10^{5}$ & 2.5$\times10^4$ & 5.0 & $1.9\times10^{5}$ & $86.0$\%\\[1.5pt]
Guo \emph{et al.} \cite{Guo:17a} & AlN & 1550 & Yes & SPDC & 5.8$\times10^{6}$ & 2$\times10^5$ & 0.7 & $8.3\times10^{6}$ & --\\[1.5pt]
Ma \emph{et al.} \cite{Ma:20} & TFLN & 1550 & No & SPDC & 2.7$\times10^{9}$ & 1.0$\times10^5$ & 1.2 & $2.3\times10^{9}$ & --\\[0.5pt]
Harper \emph{et al.} \cite{Harper:24} & TFLN & 810 & No & SPDC & 2.3$\times10^{11}$ &-- & Broadband & $3.2\times10^{6}$ & 86.0\%\\[0.5pt]
Jiao \emph{et al.} \cite{Jiao:25} & TFLN & 1560 & No & SPDC & 4.5$\times10^{10}$ & -- & Broadband & $5.0\times10^{6}$ & $\sim93\%$\\[0.5pt]
Zhao \emph{et al.} \cite{Zhao:20b} & TFLN & 1570 & No & SPDC & 4.5$\times10^{7}$ & -- & Broadband & $4.6\times10^{5}$ & 99.5\%\\[0.5pt]
Steiner \emph{et al.} \cite{Steiner:21} & AlGaAs & 1550 & No & SFWM & 2$\times10^{10}$ & 1.2$\times10^6$ & 0.10 & $2\times10^{11}$ & 97.1\%\\[1.5pt]
Kumar \emph{et al.} \cite{Kumar:19} & InP & 1550 & No & SFWM & 1.45$\times10^{8}$ & 4.2$\times10^4$ & 2.9 & $5.0\times10^{7}$ & --\\[1.5pt]
Zeng \emph{et al.} \cite{Zeng:24} & GaN & 1550 & No & SFWM & 2.09$\times10^{6}$ & 4.3$\times10^5$ & 0.29 & $7.2\times10^{6}$ & 95.5\%\\[1.8pt]
Zhao \emph{et al.} \cite{Zhao:22} & InGaP & 1550 & No & SPDC & 2.75$\times10^{10}$ & 1.1$\times10^5$ & 1.2 & $2.29\times10^{10}$ & --\\[0.5pt]
\colrule\\[-8pt]
Cao \emph{et al.} \cite{Cao:18} & PPKTP & 810 & Crystal & SPDC & 6.3$\times10^{7}$ & -- & Broadband & -- & 81.0\%\footnote{With 16-mW pump}\\[0.5pt]
Zhuang \emph{et al.} \cite{Zhuang:25} & PPLN & 1560 & Crystal & SPDC & 2.4$\times10^{10}$ & -- & Broadband & 6.4$\times 10^6$ & 99.2\%\\[0.5pt]
\end{tabular}
\end{ruledtabular}
\label{tab:S1}
\end{table*}

\clearpage
\twocolumngrid
\bibliographystyle{apsrev4-1}
\bibliography{bibliography, bib_add}

\end{document}


\title{Supplementary Materials for: \\
A scalable near-visible integrated photon-pair source for satellite quantum science}

\author{Yi-Han Luo}
\thanks{These authors contributed equally to this work.}
\affiliation{International Quantum Academy, Shenzhen 518048, China}

\author{Yuan Chen}
\thanks{These authors contributed equally to this work.}
\affiliation{International Quantum Academy, Shenzhen 518048, China}

\author{Ruiyang Chen}
\thanks{These authors contributed equally to this work.}
\affiliation{International Quantum Academy, Shenzhen 518048, China}
\affiliation{Hefei National Research Center for Physical Sciences at the Microscale and School of Physical Sciences, University of Science and Technology of China, Hefei 230026, China}

\author{Zeying Zhong}
\affiliation{International Quantum Academy, Shenzhen 518048, China}
\affiliation{Southern University of Science and Technology, Shenzhen 518055, China}

\author{Sicheng Zeng}
\affiliation{International Quantum Academy, Shenzhen 518048, China}
\affiliation{Southern University of Science and Technology, Shenzhen 518055, China}

\author{Baoqi Shi}
\affiliation{International Quantum Academy, Shenzhen 518048, China}
\affiliation{Hefei National Research Center for Physical Sciences at the Microscale and School of Physical Sciences, University of Science and Technology of China, Hefei 230026, China}

\author{Sanli Huang}
\affiliation{International Quantum Academy, Shenzhen 518048, China}
\affiliation{Hefei National Research Center for Physical Sciences at the Microscale and School of Physical Sciences, University of Science and Technology of China, Hefei 230026, China}

\author{Chen Shen}
\affiliation{International Quantum Academy, Shenzhen 518048, China}
\affiliation{Qaleido Photonics, Shenzhen 518048, China}

\author{Hui-Nan Wu}
\affiliation{Hefei National Research Center for Physical Sciences at the Microscale and School of Physical Sciences, University of Science and Technology of China, Hefei 230026, China}
\affiliation{Hefei National Laboratory, University of Science and Technology of China, Hefei 230088, China}
\affiliation{Shanghai Research Center for Quantum Sciences and CAS Center for Excellence in Quantum Information and Quantum Physics, University of Science and Technology of China, Shanghai 201315, China} 

\author{Yuan Cao}
\affiliation{Hefei National Research Center for Physical Sciences at the Microscale and School of Physical Sciences, University of Science and Technology of China, Hefei 230026, China}
\affiliation{Hefei National Laboratory, University of Science and Technology of China, Hefei 230088, China}
\affiliation{Shanghai Research Center for Quantum Sciences and CAS Center for Excellence in Quantum Information and Quantum Physics, University of Science and Technology of China, Shanghai 201315, China} 

\author{Junqiu Liu}
\email{liujq@iqasz.cn}
\affiliation{International Quantum Academy, Shenzhen 518048, China}
\affiliation{Hefei National Laboratory, University of Science and Technology of China, Hefei 230088, China}

\maketitle
\tableofcontents
\clearpage

\section{Theoretical analysis and derivation of photon\\ linewidth, PGR, and photon-pair flux}
\vspace{0.2cm}

The intracavity photon-pair generation rate (PGR) and the photon linewidth for cavity-enhanced spontaneous parametric down-conversion (SPDC) were originally derived for Fabry–P\'erot cavity configurations \cite{Ou:1999}. 
In this work, however, we generate photon pairs via cavity-enhanced spontaneous four-wave mixing (SFWM) within a microresonator. 
In this section, we theoretically derive the PGR and the photon linewidth specifically for the microresonator architecture, consisting of a closed-loop ring waveguide coupled to a bus waveguide for input and output coupling. 
The theoretical formulation follows Refs. \cite{Ou:1999, Walls2025}.

The microresonator is modelled as a quantum oscillator exhibiting inherent Kerr nonlinearity.
Upon incidence of the CW pump, photon pairs are generated in the signal (s) and idler (i) modes from the pump mode.
The interaction Hamiltonian is given by
\begin{equation}
    H_\mathrm{sys} = i\hbar\gamma\left[
	\epsilon^2 a_\mathrm{i}^\dagger a_\mathrm{s}^\dagger - \epsilon^{*2} a_\mathrm{i} a_\mathrm{s}
\right], 
\label{eqn:hamiltonian}
\end{equation}
where $a^\dagger_\mathrm{s(i)}$ and $a_\mathrm{s(i)}$ denote the creation and annihilation operators for the signal (idler) mode, respectively. 
The term $\epsilon$ represents the intracavity amplitude of the pump field, treated as a coherent state $|\epsilon\rangle$, and $\gamma=\hbar\omega^2 n_2v_g^2/(cV_\mathrm{eff})$ is the nonlinear coupling coefficient \cite{Vernon:15}. 
In the definition of $\gamma$, 
$\omega/2\pi$ is the photon's frequency (assuming frequencies for the pump, signal and idler photons are approximately the same), 
$n_2$ is the nonlinear refractive index, 
$v_g$ is the group velocity, 
$c$ is the speed of light in vacuum, 
and $V_\mathrm{eff}=A_\mathrm{eff}L$ is the effective mode volume (where $L$ is the microresonator circumference and $A_\mathrm{eff}$ is the effective mode area). 
The system is coupled to an external input-output channel and an intrinsic loss channel with coupling rates $\kappa_\mathrm{ex}$ and $\kappa_0$, respectively. 
The total decay rate is $\kappa = \kappa_\mathrm{ex} + \kappa_0$, where $\kappa/2\pi$ corresponds to the FWHM of the microresonator's resonance linewidth.

To describe the system dynamics, we employ the Heisenberg-Langevin equations, yielding
\begin{equation}
\frac{\mathrm{d}a_\mathrm{s(i)}}{\mathrm{d}t} = \gamma\epsilon^2 a_\mathrm{i(s)}^\dagger - \frac{\kappa_\mathrm{s(i)}}{2}a_\mathrm{s(i)} + \sqrt{\kappa_\mathrm{s(i), \mathrm{ex}}} a_\mathrm{s(i), \mathrm{in}} + \sqrt{\kappa_\mathrm{s(i), 0}} a_\mathrm{s(i), \mathrm{loss}}, 
\label{eq:final}
\end{equation}
where the subscript s (i) denotes the operator corresponding to the signal (idler) mode, 
$a_\mathrm{\mathrm{in}}$ and $a_\mathrm{\mathrm{loss}}$ correspond to the vacuum noises coupled to the input-output and the intrinsic loss channels. 
It is crucial to emphasize that the coupling to these vacuum fields cannot be disregarded in the quantum regime. 

Equations (\ref{eq:final}) are subsequently transformed into the frequency domain and solved for the steady state.
By invoking the basic commutation relation $[a, a^\dagger] = 1$ and the input-output relation $a_\mathrm{out} = a_\mathrm{in} - \sqrt{\kappa_\mathrm{ex}} a$, the spectral densities of the generated signal and idler photons emerging from the output port are derived as
\begin{equation}
S_\mathrm{s(i)}^\mathrm{out}(\omega) = \langle a_\mathrm{s(i), \mathrm{out}}^\dagger a_\mathrm{s(i), \mathrm{out}}\rangle (\omega) = \frac{16\kappa_\mathrm{s(i), \mathrm{ex}}\kappa_\mathrm{i(s)} \gamma^2|\epsilon|^4}{(\kappa_\mathrm{s}^2+4\omega^2)(\kappa_\mathrm{i}^2+4\omega^2)},
\label{eq:spectral}
\end{equation}
which indicate that the signal and idler photons exhibit identical spectral profiles.
The FWHM of this spectral distribution, denoted as $\Delta\nu$, is calculated as
\begin{equation}
  \Delta\nu = \frac{1}{2\pi\sqrt{2}}\sqrt{\sqrt{(\kappa_\mathrm{s}^2+\kappa_\mathrm{i}^2)^2 + 4\kappa_\mathrm{s}^2\kappa_\mathrm{i}^2}-(\kappa_\mathrm{s}^2+\kappa_\mathrm{i}^2)}. 
  \label{eqn:linewidth}
\end{equation}
In the case where the decay rates are identical (i.e., $\kappa_\mathrm{s} = \kappa_\mathrm{i} = \kappa$), the expression simplifies to $\Delta \nu \approx 0.64\kappa/2\pi$, which is consistent with the theoretical conclusion reported in Ref. \cite{Ou:1999}. 
For our experimental parameters, where $\kappa_\mathrm{s}/2\pi=605$ MHz and  $\kappa_\mathrm{i}/2\pi=514$ MHz, the calculated linewidth is $\Delta\nu=357$ MHz. 

To derive the PGR, we first assume that $\kappa_0$ and $\kappa_\mathrm{ex}$ are uniform across all interacting modes for simplicity. 
Integrating Eq. (\ref{eq:spectral}) over the frequency domain $\omega$ yields the photon flux at the output port
\begin{equation}
    N_\mathrm{out} = \frac{2\kappa_\mathrm{ex}\gamma^2|\epsilon|^4}{\kappa^2}. 
\end{equation}
Given the photon extraction efficiency $\eta = \kappa_\mathrm{ex}/\kappa$, the intracavity photon-pair flux $\alpha$ can be expressed as
\begin{equation}
    \alpha = \frac{2\gamma^2|\epsilon|^4}{\kappa}.
\end{equation}
Furthermore, by substituting the relation for the intracavity pump photon number $|\epsilon|^2 = (4\kappa_\mathrm{ex}/\kappa^2)\cdot(P/\hbar\omega)$, where $P$ is the on-chip pump power (i.e. the pump power in the bus waveguide), we obtain
\begin{equation}
    \alpha = \frac{32 \kappa_\mathrm{ex}^2}{\kappa^5} \left(\frac{\gamma P}{\hbar\omega}\right)^2 =\frac{32 \kappa_\mathrm{ex}^2}{\kappa^5} \left(\frac{\hbar\omega^2 n_2v_g^2}{cV_\mathrm{eff}}\right)^2 \left(\frac{P}{\hbar\omega}\right)^2  = \mathrm{PGR}\times P^2. 
    \label{eqn:generation}
\end{equation}
With $V_\mathrm{eff}=A_\mathrm{eff}L$, the scaling law of Eq. (2) in the main text is readily derived.
An estimation is made using Eq. (\ref{eqn:generation}).
Using the experimental parameters associated with the TE$_{20}$ mode as
$\omega/2\pi=384\times10^{12}$ Hz, 
$n_2=3.0\times10^{-19}$ m$^2$/W (accounting for the slight increase relative to that of telecommunication bands), 
$v_g = 1.5\times10^8$ m/s, 
$V_\mathrm{eff}=2\pi RA_\mathrm{eff}$ ($A_\mathrm{eff}=0.5$ $\mu$m$^2$, 
$R = 41.6$ $\mu$m), 
$P=10^{-3}$ W, 
$\kappa_\mathrm{ex}/2\pi=180$ MHz, 
$\kappa/2\pi=530$ MHz, 
we obtain a theoretical value of $\mathrm{PGR} \approx 1.7\times10^7$ pairs/s/mW$^2$.
This result shows excellent agreement with the experimental result for the $\mathrm{TE}_{20}$ mode reported in the main text.

By combining Eq. (\ref{eqn:generation}) with the photon extraction efficiency $\eta = \kappa_\mathrm{ex}/\kappa$, the heralding rate $n_1$ (defined as the single-photon detection rate in one channel) is given by
\begin{equation}
    n_1 = \frac{32 \kappa_\mathrm{ex}^3}{\kappa^6} \left(\frac{\hbar\omega^2 n_2v_g^2}{cV_\mathrm{eff}}\right)^2 \left(\frac{P}{\hbar\omega}\right)^2.
    \label{eqn:heralding}
\end{equation}
Similarly, the photon-pair flux coupled into the bus waveguide, $\alpha_\mathrm{bus}$, is expressed as
\begin{equation}
    \alpha_\mathrm{bus} = \frac{32 \kappa_\mathrm{ex}^4}{\kappa^7} \left(\frac{\hbar\omega^2 n_2v_g^2}{cV_\mathrm{eff}}\right)^2 \left(\frac{P}{\hbar\omega}\right)^2.
    \label{eqn:flux_bus}
\end{equation}
For a specific microresonator design operating at a fixed frequency and fabricated with a stable process, the geometric and material parameters in Eqs. (\ref{eqn:generation}--\ref{eqn:flux_bus}) remain constant.
The primary degree of freedom for optimization is $\kappa_\mathrm{ex}$, which can be precisely tuned by modifying the coupling region geometry. 
Consequently, the maxima for $\alpha$, $n_1$, and $\alpha_\mathrm{bus}$ are achieved when $\kappa_\mathrm{ex} = 2\kappa_0/3$, $\kappa_\mathrm{ex} = \kappa_0$, and $\kappa_\mathrm{ex} = 4\kappa_0/3$, respectively.

In applications where maximizing the photon-pair flux is paramount, it is essential to consider the flux at the parametric oscillation threshold power $P_\mathrm{th}$. 
According to Ref. \cite{Pu:16}, the threshold power is defined as
$$
P_\mathrm{th} = A\cdot\frac{cV_\mathrm{eff}}{\omega n_2 v_g^2}\cdot\frac{\kappa^3}{\kappa_\mathrm{ex}}, 
$$
where $A$ is a dimensionless coefficient that does not influence the fundamental scaling behavior investigated here. 
By substituting $P_\mathrm{th}$ into Eq. (\ref{eqn:generation}), we obtain the maximum achievable photon-pair flux at the threshold
\begin{equation}
    \alpha^\mathrm{max} = \mathrm{PGR}\times P_\mathrm{th}^2 = 32A^2 \kappa. 
\end{equation}
Similarly, the maxima for $n_1$ and $\alpha_\mathrm{bus}$ are $32A^2\kappa_\mathrm{ex}$ and $32A^2\kappa_\mathrm{ex}^2/\kappa$, respectively. 

These results imply that for a fixed $\kappa_0$, increasing $\kappa_\mathrm{ex}$ simultaneously increases $\alpha$, $n_1$, and $\alpha_\mathrm{bus}$ at the threshold. 
However, in practical deployments, a high photon-pair flux often entails trade-offs, including degraded single-photon purity (higher multi-photon probability), reduced two-photon interference visibility, increased spectral linewidth, and higher required pump powers. 
These competing factors must be carefully balanced to meet the specific requirements in quantum applications.

\clearpage

\section{Device fabrication process}
\vspace{0.2cm}

\begin{figure*}[h!]
\centering
\includegraphics[width=0.95\textwidth]{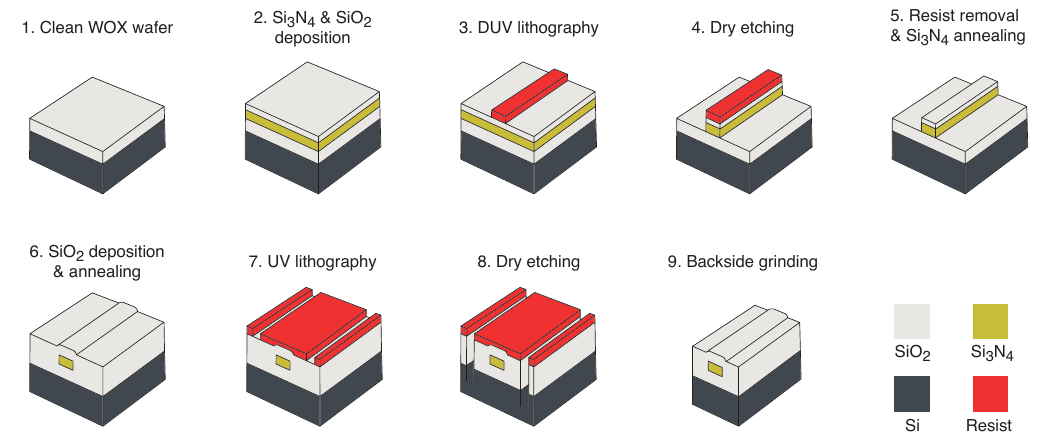}
\caption{
\textbf{The DUV subtractive process flow of 6-inch-wafer Si$_3$N$_4$ foundry fabrication.}
WOX, thermal wet oxide (SiO$_2$). }
\label{Fig:S0}
\end{figure*}

The Si$_3$N$_4$ photonic chips are fabricated using an optimized deep-ultraviolet (DUV) subtractive process on 6-inch wafers \cite{Ye:23, Sun:25}, performed in our CMOS foundry.
The process flow is illustrated in Fig. \ref{Fig:S0}. 
First, a 300-nm-thick Si$_3$N$_4$ film is deposited on a clean thermal wet SiO$_2$ substrate via low-pressure chemical vapor deposition (LPCVD).
Subsequently, an SiO$_2$ film is deposited on the Si$_3$N$_4$ as an etch hardmask, again via LPCVD.
Next, DUV stepper lithography is performed, followed by dry etching to transfer the pattern from the DUV photoresist to the SiO$_2$ hardmask, and then to the Si$_3$N$_4$ layer.
The dry etching uses etchants comprising CHF$_3$ and O$_2$ to create ultra-smooth and vertical etched sidewalls, critical for minimizing optical losses in waveguides.
Afterward, the photoresist is removed, and thermal annealing of the entire wafer is applied under a nitrogen atmosphere at 1200 $^\circ$C.
Then a 3-$\mu$m-thick SiO$_2$ top cladding layer is deposited on the wafer, followed by another thermal annealing at 1200 $^\circ$C.
Finally, UV photolithography and deep dry etching are performed to create smooth chip facets facilitating fiber coupling. 
The wafer is separated into individual chips through backside grinding or dicing.

The optimization of the LPCVD Si$_3$N$_4$ process is particularly critical for near-visible photon-pair generation, where material-induced background noise can severely degrade the coincidence-to-accidental ratio (CAR). 
In our fabrication, the LPCVD precursor ratio of ammonia (NH$_3$) to dichlorosilane (SiCl$_2$H$_2$) is precisely tuned between 7:1 and 10:1 to ensure a stoichiometric or slightly nitrogen-rich (N-rich) Si$_3$N$_4$ composition. 
Unlike films produced via plasma-enhanced chemical vapor deposition (PECVD) or silicon-rich nitride variants \cite{Kruckel:17}---which often contain excess silicon clusters and high hydrogen content that act as color centers---LPCVD-grown stoichiometric films suppress this significant fluorescence typically emitted under near-visible pumping.
Consequently, these stoichiometric films not only reduce linear propagation losses but also minimize fluorescence-induced background noise. 
As a result, the photon-pair CAR is significantly enhanced, increasing from 3 in previous work \cite{Zhao:20a} to more than 1000 in our work.


\clearpage

\section{Device characterization}
\vspace{0.2cm}

\subsection{Characterization setup}

\begin{figure*}[b!]
\centering
\includegraphics[width=\textwidth]{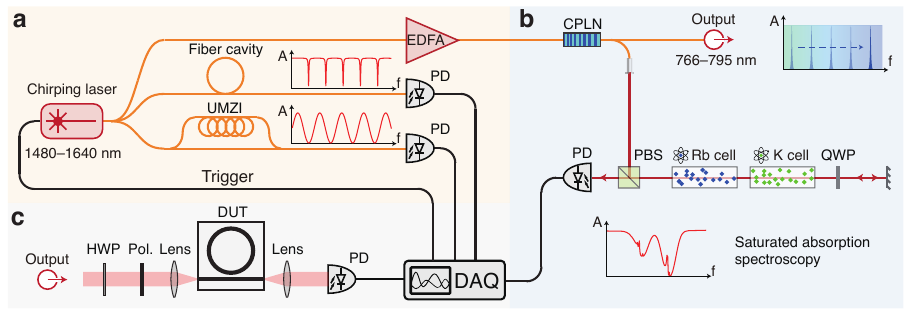}
\caption{
\textbf{Experimental setup for microresonator characterization in the near-visible spectral range.}
\textbf{a}, 
The NIR laser source and its relative frequency calibration. 
An ECDL is split into three branches, with two directed to a fiber cavity and a UMZI to provide a high-precision, on-the-fly frequency ruler and interpolation.
\textbf{b}, 
Frequency doubling and the absolute frequency reference. 
The NIR laser is amplified by an EDFA and frequency-doubled via a CPLN waveguide to the 766--795 nm range. 
The resulting SH light is used for absolute frequency calibration using saturated absorption spectroscopy with Rb and K vapor cells.
\textbf{c}, 
Configuration for the DUT. 
All transmission signals---from the DUT, atomic references, and frequency rulers---are synchronized and acquired by a DAQ module to reconstruct the high-resolution transmission spectrum.
}
\label{Fig:S1}
\end{figure*}

The experimental setup employed for characterizing microresonator loss and dispersion is illustrated in Fig. \ref{Fig:S1}. 
This configuration is based on the method developed in Ref. \cite{Shi:25}. 
As depicted in Fig. \ref{Fig:S1}a, the system utilizes a widely tunable, mode-hop-free, near-infrared (NIR), external-cavity diode laser (ECDL, Santec TSL series) capable of continuous wavelength sweeping. 
The laser output is split into three distinct paths. 
The optical signal in the top branch is amplified by an EDFA and subsequently coupled into a chirped periodically poled lithium niobate (CPLN) waveguide. 
This process generates broadband second-harmonic (SH) light in the near-visible spectral range (766--795 nm). 
To accommodate the operational bandwidths of both the EDFA and the CPLN, the tunable laser is operated within the 1532--1590 nm range. 
As shown in Fig. \ref{Fig:S1}b, the generated SH light is further split into two paths. 
The lower path is directed to an atomic reference unit comprising two vapor cells containing rubidium (Rb) and potassium (K), respectively. 
By employing saturated absorption spectroscopy, the hyperfine transitions of these alkali atoms serve as absolute frequency references. 
The upper path is routed to the device under test (DUT), as illustrated in Fig. \ref{Fig:S1}c.

The lower two paths are directed toward a fiber cavity and an unbalanced Mach-Zehnder interferometer (UMZI), respectively.
The fiber cavity's FSR is pre-calibrated using a sideband modulation technique described in Ref. \cite{Luo:24}. 
Consequently, as the NIR laser sweeps, the transmission spectrum of the fiber cavity---containing an FSR-calibrated resonance grid---serves as a precise time-domain frequency ruler. 
This allows for on-the-fly calibration of the relative frequency detuning with respect to the initial or an intermediate reference frequency. 
The UMZI is utilized to perform frequency interpolation between discrete resonant frequencies of the fiber cavity.

The transmission signals from the fiber cavity, the UMZI, the atomic references, and the DUT are synchronized with the trigger signal of the tunable laser. 
These signals are simultaneously acquired by a data acquisition (DAQ) module to reconstruct the calibrated transmission spectrum of the DUT (the microresonator) across the 766--795 nm spectral range.

The resonant frequencies are extracted from the microresonator's transmission spectrum using a peak-searching algorithm. 
These frequencies are subsequently utilized to calculate the FSR and the integrated dispersion ($D_\mathrm{int}$) profile as detailed in the main text. 
For each resonance, we fit the profile using the formula \cite{Li:13}
\begin{equation}
T = \left|1 - \frac{\kappa_\mathrm{ex}\left[i(\omega-\omega_0)+(\kappa_0+\kappa_\mathrm{ex})/2\right]}{\left[i(\omega-\omega_0)+(\kappa_0+\kappa_\mathrm{ex})/2\right]^2+(k_\mathrm{r}+ik_\mathrm{i})^2/4} \right|^2, 
\label{eqn:resonant}
\end{equation}
where $k_\mathrm{r}$ and $k_\mathrm{i}$ represent the real and imaginary parts of the complex coupling coefficient between the clockwise and counter-clockwise modes, respectively. 
Due to the mathematical symmetry between $\kappa_0$ and $\kappa_\mathrm{ex}$ in Eq. (\ref{eqn:resonant}), it is impossible to distinguish these two decay rates through a single-peak fit alone. 
However, observing that the resonance dips in Fig. 3a deepen at longer wavelengths, we infer that the microresonator operates in the under-coupled regime ($\kappa_\mathrm{ex} < \kappa_0$) within this spectral range. 

\subsection{Higher-order mode identification}

To verify that the resonances utilized for near-visible photon-pair generation correspond to the TE$_{20}$ mode, we identify the mode order by comparing the experimentally measured FSR with numerical simulation. 
Given the small microresonator radius (41.6 $\mu$m), bending-induced effects on the mode profile and dispersion cannot be neglected. 
Consequently, we solve for the frequency-dependent effective refractive index $n_\mathrm{eff}(f)$ under cylindrical coordinates $(r, \phi, z)$ using Finite Element Method (FEM).
The ansatz $\mathbf{E}(r, z, \phi) = \mathbf{E}(r, z)e^{i\xi k_0\phi}$ is employed, where $k_0$ is the wave vector in vacuum, and $\xi$ is the eigenvalue to be determined. 
Since $\xi = n_\mathrm{eff} R$, the effective refractive index is given by $n_\mathrm{eff} = \xi/R$. 
Here, $R$ represents the effective mode radius, calculated as the $r$-coordinate weighted by the $\phi$-component of the Poynting vector derived from the field distribution $\mathbf{E}(r, z, \phi)$.

\begin{figure*}[b!]
\centering
\includegraphics[width=0.95\textwidth]{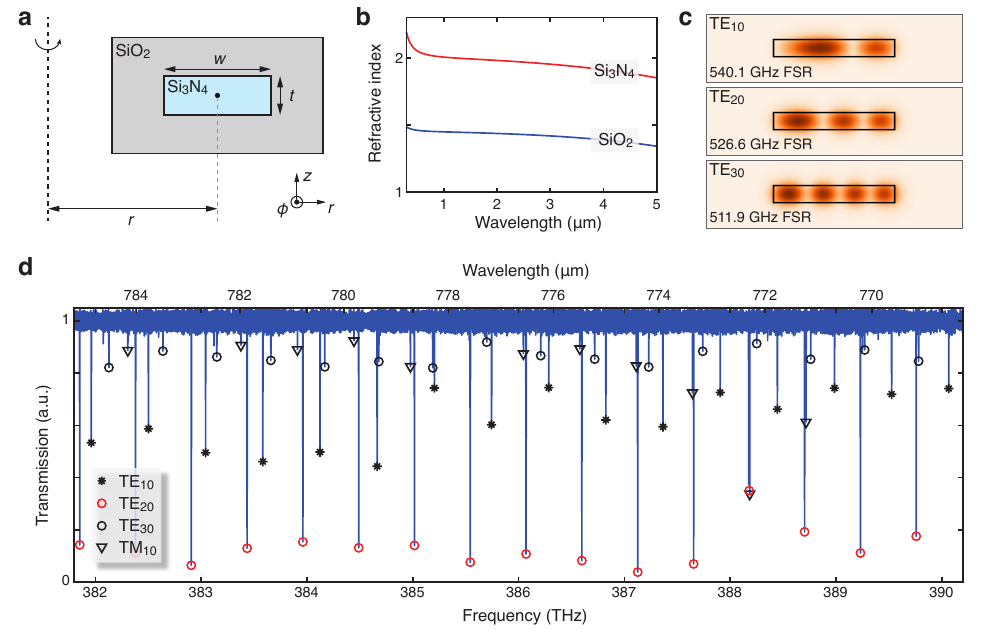}
\caption{
\textbf{Identification of higher-order modes. }
\textbf{a}, 
Geometry of the Si$_3$N$_4$ waveguide used for mode simulations.
\textbf{b}, 
Material refractive indices for Si$_3$N$_4$ and SiO$_2$ derived from Sellmeier fits of experimental data.
\textbf{c}, 
Simulated mode profiles for the TE$_{10}$, TE$_{20}$, and TE$_{30}$ modes, with their respective calculated FSR values.
\textbf{d}, 
Measured transmission spectrum of a microresonator with its bus waveguide width intentionally detuned from the TE$_{20}$ phase-matching condition, showing the excitation of multiple mode families.
}
\label{Fig:S2}
\end{figure*}

Specifically, as illustrated in Fig. \ref{Fig:S2}a, the simulation assumes a microresonator radius of $r=41.6$ $\mu$m, a waveguide width of $w=2200$ nm, and a thickness of $t=300$ nm. 
The Sellmeier equations for the constituent materials, fitted from experimental data spanning 0.3 to 5.0 $\mu$m within our fabrication process, are
\begin{subequations}
    \begin{align}
        n_\mathrm{{Si_3N_4}}^2 &= 1 + \frac{1.73783 \lambda^2}{\lambda^2 - 0.00745}
        + \frac{1.25000\lambda^2}{\lambda^2 - 0.03128}
        + \frac{2.68390\lambda^2}{\lambda^2 - 143.75857},\\
        n_\mathrm{{SiO_2}}^2 &= 1 + \frac{0.69652\lambda^2}{\lambda^2 - 0.00429}
        + \frac{0.40321\lambda^2}{\lambda^2 - 0.01401}
        + \frac{0.85807\lambda^2}{\lambda^2 - 96.57021}, 
    \end{align}
\end{subequations}
where the wavelength $\lambda$ is in units of micrometers. 
The resulting refractive index profiles are plotted in Fig. \ref{Fig:S2}b.

Once $n_\mathrm{eff}(f)$ is obtained, the propagation constant is calculated as $\beta(\omega) = n_\mathrm{eff}(f) \omega/c$, where $\omega$ is the angular frequency and $c$ is the speed of light. 
The group delay per unit length, $\beta_1 = \mathrm{d}\beta/\mathrm{d}\omega$, is extracted via a polynomial fit to the simulated dispersion data. 
Finally, the FSR of the specific mode is calculated as $f_\mathrm{FSR} = 1/(\beta_1 R)$. 
Figure \ref{Fig:S2}c displays the simulated mode profiles for the TE$_{10}$, TE$_{20}$, and TE$_{30}$ modes, with calculated FSR ($D_1/2\pi$) values of 540.1, 526.6 and 511.9 GHz, respectively.

To experimentally identify the mode families, we characterize a microresonator with its bus waveguide width intentionally detuned from the TE$_{20}$ phase-matching condition, thereby allowing for simultaneous excitation of multiple mode families. 
In the experiment, we resolve three distinct sets of resonances with measured FSR values of 540.6, 527.4 and 510.3 GHz. 
By comparing these measured values with simulated results, we unambiguously assign the resonance set with 527.4 GHz FSR to the TE$_{20}$ mode. 
Additionally, an observed mode with 534.4 GHz FSR agrees with the simulated TM$_{10}$ mode (532.8 GHz). 
The excitation of the TM mode is attributed to the bus waveguide's near-unity aspect ratio in the coupling region, which facilitates cross-polarization coupling \cite{Kordts:16}.

\clearpage

\section{Experimental setup}
\vspace{0.2cm}

\begin{figure*}[h!]
\centering
\includegraphics[width=1\textwidth]{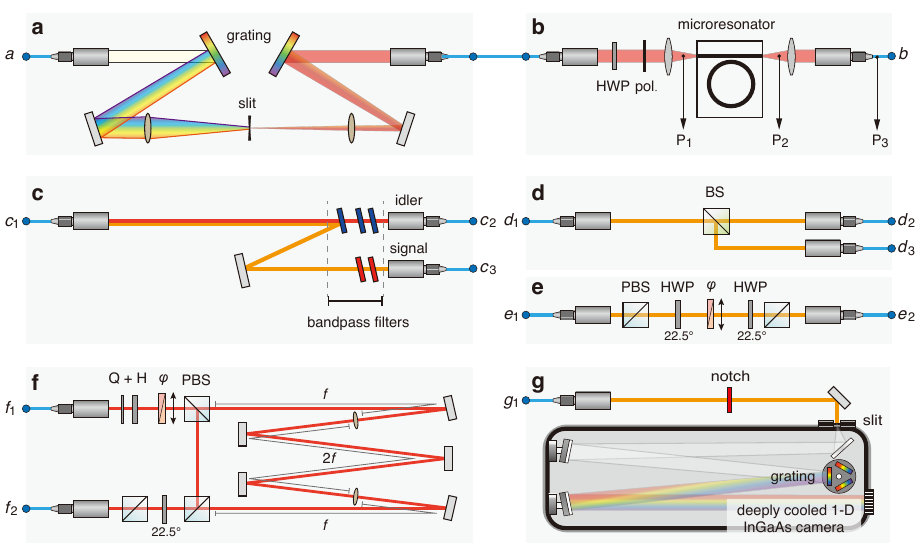}
\caption{
\textbf{Modular experimental configuration for photon-pair performance characterization.}
The setup is partitioned into distinct functional modules interconnected via single-mode fibers and flanges, allowing for measurement-specific reconfiguration.
\textbf{a}, 
Grating-based 4-$f$ spectral cleaning system to suppress pump sideband noise.
\textbf{b}, 
Chip-coupling interface via edge coupling.
\textbf{c}, 
Multi-stage filtering system for high-extinction pump rejection ($>100$ dB) and spatial separation of signal and idler photons.
\textbf{d}, 
HBT interferometer for heralded second-order correlation $g^{(2)}(0)$ measurements.
\textbf{e}, 
Polarization interferometer for heralded single-photon interference.
\textbf{f}, 
Folded Franson interferometer with a 4-$f$ mode-matching system to resolve energy-time entanglement interference fringes.
\textbf{g}, 
Grating spectrometer equipped with a deeply cooled InGaAs camera for high-resolution emission spectral analysis.
}
\label{Fig:S3}
\end{figure*}

The detailed experimental configuration employed to characterize the performance of the photon-pair source is illustrated in Fig. \ref{Fig:S3}. 
The setup is designed with a modular architecture, partitioned into several distinct functional subsections. 
Each module is interfaced via single-mode fibers for both input and output coupling, with interconnections established through standard fiber flanges. 
This modular approach provides exceptional flexibility, allowing for diverse characterizations to be executed by reconfiguring the modules connection in various sequences.

For photon-pair generation, the CW pump laser first passes through a 4-$f$ optical system designed to filter out sideband noise from the laser source. 
As illustrated in Fig. \ref{Fig:S3}a, this system consists of two identical lenses with a focal length of $f=200$ mm.
The incident laser beam is initially diffracted by a first grating (featuring a blazing wavelength of 750 nm and a groove density of 1200 g/mm), ensuring that different frequency components are distributed along different spatial directions. 
A slit positioned precisely in the Fourier plane serves as a bandpass filter to isolate the desired frequency components. 
The filtered pump light is subsequently frequency-recombined by a second grating. 
To ensure optimal performance, the central wavelength of this passband is carefully adjusted to match the resonant frequency of the pump mode in the microresonator used for photon-pair generation.

Subsequently, the laser is directed to the microresonator as illustrated in Fig. \ref{Fig:S3}b. 
To achieve optimal coupling, we align the incident beam's polarization to the TE$_{00}$/TM$_{00}$ mode of the bus waveguide using a linear polarizer.
A preceding half-wave plate (HWP) is employed to rotate the polarization incident on the polarizer, thereby facilitating precise and continuous adjustment of the on-chip pump power.
The polarized light is coupled into and out of the chip using a pair of aspherical lenses (Thorlabs C392TME-B). 
To characterize the coupling performance of the input and output facets, we measure the optical power at three specific locations, i.e. $P_1$ (incident power in free space), $P_2$ (transmitted power in free space), and $P_3$ (power collected into the fiber). 
By detuning the laser from the microresonator resonance to avoid cavity absorption, the input coupling efficiency is determined as $\eta_{\mathrm{in}}=P_2/P_1$, while the overall system throughput is defined as $\eta_{\mathrm{in}}\eta_{\mathrm{out}}=P_3/P_1$. 
Through this procedure, we obtain measured coupling values of $\eta_{\mathrm{in}}\approx0.33$ and $\eta_{\mathrm{out}}\approx0.50$.

To spatially separate the generated signal and idler photons, a multi-stage bandpass filtering system is employed. 
As illustrated in Fig. \ref{Fig:S3}c, the signal and idler photons undergo initial separation at the first filter stage through transmission and reflection.
Subsequently, a series of filters are utilized to further suppress frequencies external to the signal and idler bands, specifically the residual pump laser light that remains collinearly transmitted. 
To verify the filtering effectiveness, we experimentally detune the pump frequency slightly from the microresonator resonance to inhibit photon-pair generation. 
Filters are then one-by-one inserted into the idler and signal paths until the photon count rates registered by the APDs decrease to their intrinsic dark count levels. 
This protocol confirms that the pump laser is effectively excluded. 

The functional modules illustrated in Fig. \ref{Fig:S3}d, e are independently utilized for the characterization of heralded single-photon properties, specifically the heralded second-order correlation function $g^{(2)}(0)$ and single-photon interference visibility.
To measure $g^{(2)}(0)$ and to verify single-photon purity, the heralded signal photons are directed into a 50:50 beam splitter (BS), which splits the signal into two symmetric paths for coincidence detection.
In Fig. \ref{Fig:S3}e, to quantify single-photon interference visibility, the input photons are first prepared in the superposition state $(|H\rangle+|V\rangle)/\sqrt{2}$ using a combination of polarizing beam splitter (PBS) and an HWP. 
A relative phase $\varphi$ is subsequently introduced between the horizontal ($H$) and vertical ($V$) components via a specialized phase shifter.
This phase shifter consists of two angled quartz plates glued together with their optical axes oriented perpendicular to each other. 
By translating phase shifter, the phase is precisely tuned.
The photons are projected onto the initial state again to resolve the interference fringes.

To resolve the two-photon interference visibility of the energy-time entanglement, a folded Franson interferometer is employed in Fig. \ref{Fig:S3}f. 
The polarization of incident photons is precisely adjusted using a combination of a quarter-wave plate (QWP) and an HWP---denoted as ``Q+P'' in the schematic---to ensure the flux is equally split by the subsequent PBS.
Following the first PBS, the photons are distributed into two arms with a length difference of 2 m, corresponding to a temporal delay of approximately 6.6 ns. 
This delay is designed to be significantly larger than the two-photon correlation histogram FWHM (0.7 ns for TE$_{20}$ and 1.2 ns for TM$_{10}$), ensuring that the single-photon interference is prohibited. 
The two paths are recombined at a second PBS and projected onto the superposition state $(|H\rangle+|V\rangle)/\sqrt{2}$ to observe the two-photon interference.
To control the quantum interference, the same birefringent phase shifter utilized in Fig. \ref{Fig:S3}e is inserted before the first PBS to adjust the relative phase between the two arms. 
Notably, because the folded Franson interferometer possesses an unbalanced configuration where photons propagate through vastly different lengths, spatial mode mismatch can occur. 
To ensure the mode profiles from both arms are identical upon recombination at the second PBS and subsequent collection by the collimator, a 4-$f$ optical system consisting of two lenses with a 400-mm focal length is integrated. 
This 4-$f$ system ensures that the spatial mode transformation corresponding to the long arm approximates the identity matrix, thereby maintaining high interference visibility.

Additionally, we analyze the noise properties of the photon pairs by directly measuring the spectral distribution of the emission from the microresonator. 
As illustrated in Fig. \ref{Fig:S3}g, the emitted photons first pass through a notch filter designed to exclude the pump laser with an extinction ratio exceeding 100 dB.
The filtered signal is subsequently directed to a grating spectrometer (Princeton Instrument HRS-750) equipped with a deeply cooled 1-D InGaAs camera (Princeton Instrument PyLoN-IR 1.7).
It should be noted that although this camera is primarily optimized for the NIR spectrum (800--1700 nm), it maintains a quantum efficiency of approximately 20\% at 780 nm. 
This sensitivity is sufficient for characterizing the noise spectra in the near-visible regime of our source.

In Fig. \ref{Fig:S3}, all ports of the individual modules are labelled, and their specific interconnections are defined separately according to the required measurement:
\begin{itemize}[nosep]
    \item Spectral brightness and CAR: pump $\rightarrow~a$, $b\rightarrow c_1$, $c_2\&c_3\rightarrow$ APDs.
    \item Heralded second-order correlation: pump $\rightarrow~a$, $b\rightarrow c_1$, $c_3\rightarrow d_1$, $c_2\&d_2\&d_3\rightarrow$ APDs. 
     \item Heralded single-photon interference: pump $\rightarrow~a$, $b\rightarrow c_1$, $c_3\rightarrow e_1$, $c_2\&e_2\rightarrow$ APDs. 
     \item Two-photon entanglement interference: pump $\rightarrow~a$, $b\rightarrow f_1$, $f_2\rightarrow c_1$, $c_2\&c_3\rightarrow$ APDs. 
     \item Emission spectral analysis: pump $\rightarrow~a$, $b\rightarrow g_1$.
\end{itemize}

\clearpage

\section{Supplementary experimental results and discussion}
\vspace{0.2cm}

\subsection{Near-visible photon-pair generation with TM$_{10}$ mode}

\begin{figure*}[b!]
\centering
\includegraphics[width=\textwidth]{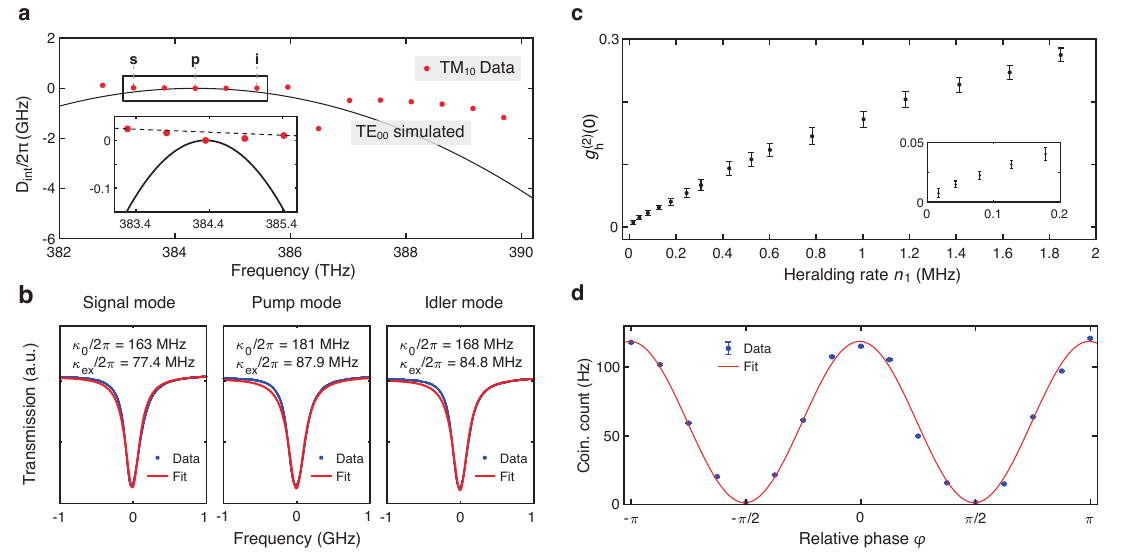}
\caption{
\textbf{Characterization of photon pairs generated from the TM$_{10}$ mode.}
\textbf{a}, 
Measured $D_\mathrm{int}/2\pi$ profile. 
The experimental data confirm the existence of local anomalous GVD of the TM$_{10}$ mode, in sharp contrast to the normal GVD exhibited by the TE$_{00}$ mode (simulated, solid curve).
Inset: Magnified view of the phase-matching window. 
The pump mode frequency lies below the arithmetic mean of the signal and idler frequencies (dashed line), satisfying the phase-matching conditions required for SFWM. 
\textbf{b}, 
Resonance profiles of the signal, pump and idler modes, overlaid with their Lorentzian fits.
The extracted $\kappa_0/2\pi$ and $\kappa_\mathrm{ex}/2\pi$ are marked for each mode.
\textbf{c}, 
Heralded single-photon purity $g^{(2)}_\mathrm{h}(0)$ as a function of heralding rate $n_1$. 
The heralded $g^{(2)}_\mathrm{h}(0)$ remains low even as $n_1$ reaches 2.1 MHz at $P=0.7$ mW, confirming robust single-photon statistics. 
Inset: Magnified view of the zoom $g^{(2)}_\mathrm{h}(0)<0.05$.
\textbf{d}, 
Measured two-photon interference fringe fitted with a sinusoidal function (red curve), yielding a visibility of $V=97.9\%$. 
Error bars represent one standard deviation derived from Poissonian counting statistics, which are however much smaller than the data point size. 
}
\label{Fig:S5}
\end{figure*}

Leveraging wafer-scale fabrication with sweeping geometric parameters, we have identified a microresonator whose $\mathrm{TM}_{10}$ mode exhibits local anomalous GVD induced by an AMX. 
We experimentally characterize the photon-pair performance based on this device.
By subtracting the FSR ($D_1/2\pi = 534.4$ GHz), the $D_\mathrm{int}/2\pi$ profile of the $\mathrm{TM}_{10}$ mode is extracted and presented in Fig. \ref{Fig:S5}a. 
We identify the target signal, pump, and idler resonances at 383.28230, 384.35063 and 385.41900 THz, respectively. 
The transmission profiles of these interacting $\mathrm{TM}_{10}$ modes are detailed in Fig. \ref{Fig:S5}b. 
Lorentzian fitting of these resonances yields total decay rates $(\kappa_0+\kappa_\mathrm{ex})/2\pi$ of 240, 269 and 253 MHz for the signal, pump, and idler modes, corresponding to loaded $Q$ values of $1.6\times10^6$, $1.4\times10^6$, and $1.5\times10^6$, respectively.

Due to the enhanced $Q$ factor, the resulting two-photon correlation histogram exhibits an FWHM of 1.2 ns. 
We further extract PGR of $a = 9.29 \times 10^7$ pairs/s/mW$^2$.
Combined with the narrow photon linewidth of $\Delta\nu = 159$ MHz, this yields a spectral brightness of $5.84 \times 10^8$ pairs/s/mW$^2$/GHz. 
The CAR reaches a peak value of $350 \pm 10$ at a pump power of $P = 24$ $\mu$W. 
Notably, this maximum CAR is lower than that obtained for the $\text{TE}_{20}$ mode presented in the main text. 
A comprehensive noise analysis for this specific mode is provided in the subsequent subsection \textbf{C}.

Figure \ref{Fig:S5}c displays the measured $g^{(2)}_\mathrm{h}(0)$ as a function of the heralding rate $n_1$. 
At $P=36$ $\mu$W and utilizing a coincidence window width of 1.2 ns, we observe $g^{(2)}_\mathrm{h}(0) = 0.0073 \pm 0.0037$ at $n_1=17$ kHz. 
Consistent with the behavior observed for the $\mathrm{TE}_{20}$ mode, even as the pump power approaches the OPO threshold power ($P_\mathrm{th} \approx 0.7$ mW), the heralding rate reaches $n_1=2.1$ MHz while $g^{(2)}_\mathrm{h}(0)$ remains below 0.3.
The two-photon interference fringe characterizing the energy-time entanglement is presented in Fig. \ref{Fig:S5}d. 
The measurement is conducted at $P=62$ $\mu$W using a coincidence window width of 4.0 ns. 
Each data point represents the coincidence count rate $n_\mathrm{cc}$ averaged over an integration time of 30 seconds. 
Without any background subtraction, the raw interference visibility is determined through a sinusoidal fit to be $V = (97.9 \pm 1.9)\%$. 

\subsection{Two-photon intereference histogram for different pump powers}

\begin{figure*}[h!]
\centering
\includegraphics[width=0.95\textwidth]{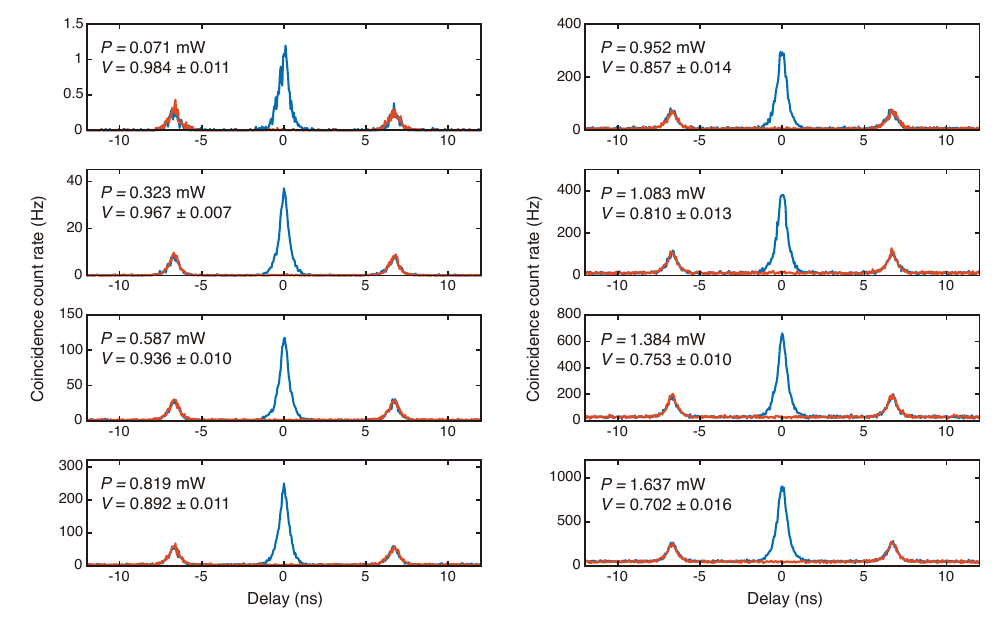}
\caption{
\textbf{Two-photon interference histogram for different pump powers. }
The three-peak two-photon interference (TPI) histogram for different pump power $P$ are displayed. 
The constructive (blue lines) and destructive (red lines) interference are outlined. 
The coincidence count rate corresponds to a bin width of 50 ps. 
The pump power $P$ varies from 0.071 to 1.637 mW, with the TPI visibility (integrated over 3.0 ns) decreases from 0.984 to 0.702. 
}
\label{Fig:S4}
\end{figure*}

In the main text, we characterize the dependence of the two-photon interference visibility $V$ on the pump power $P$, derived from the two-photon coincidence histograms. 
Figure \ref{Fig:S4} displays these histograms for varying $P$ from 0.071 to 1.637 mW. 
As $P$ increases, $V$ decreases from 0.984 to 0.702.
Notably, even at the highest $P$---where $V$ falls slightly below the CHSH violation limit ($V \approx 70.7\%$)---the central coincidence peak vanishes almost completely under destructive interference. 
This indicates that the folded Franson interferometer maintains high stability and precise phase control throughout the measurement range. 
The degradation of $V$ at higher $P$ is primarily attributed to the elevation of the noise floor, which becomes increasingly prominent as $P$ increases. 
A detailed analysis of this noise is provided in the following subsection.

\subsection{Noise analysis}

\begin{figure*}[t!]
\centering
\includegraphics[width=\textwidth]{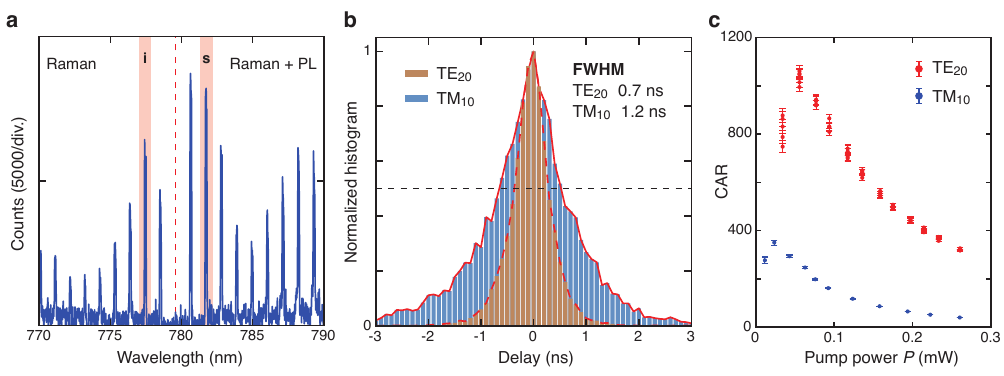}
\caption{
\textbf{Noise analysis. }
\textbf{a}, 
The photon emission spectrum from the TE$_{20}$ mode. 
The spectrum is obtained with $P=443$ $\mu$W and an integration time of 5400 seconds. 
The asymmetry of the spectrum reveals the existence of photoluminescence photons on the Stokes side (longer wavelengths). 
\textbf{b}, 
The two-photon correlation histogram of photon pairs from the TM$_{10}$ and TE$_{20}$ modes. 
The higher $Q$ factor of the TM$_{10}$ mode results in a larger FWHM. 
\textbf{c}, 
A comparison of CAR for the photon-pair sources based on the TM$_{10}$ and TE$_{20}$ modes. 
The maximum CAR of the TM$_{10}$ mode is $350\pm10$, much lower than that of the TE$_{20}$ mode. 
}
\label{Fig:S6}
\end{figure*}

We first characterize the emission spectrum of the microresonator’s $\mathrm{TE}_{20}$ mode. 
The measured spectrum, obtained with $P=443$ $\mu$W and an integration time of 5400 seconds, is presented in Fig. \ref{Fig:S6}a. 
Figure \ref{Fig:S6}a shows the asymmetric photon counts: 
the count rates at wavelengths longer than the pump (Stokes side) are notably higher than those at shorter wavelengths (anti-Stokes side). 
This spectral asymmetry reveals that the background noise consists not only of spontaneous Raman scattering---originating from the Boson peak of the amorphous medium \cite{Buchenau:86}---but also of photoluminescence (PL) from residual silicon clusters \cite{Zhao:20a}. 
Although the film deposition process via LPCVD is optimized to minimize such defects, these silicon clusters remain an unavoidable source of broadband emission.
This noise, distributed uniformly in time, is efficiently suppressed through coincidence gating, thereby contributing to a small accidental coincidence floor. 
CAR acts as a key metric to reflect the noise suppression.
While the background noise photons only slightly constrain the performance of the heralded single photon and energy-time entangled sources, the CAR remains a critical metric for quantifying the impact of this noise. 
Reported CAR values in microresonators range from several hundreds to over 12,000 \cite{Ma:17}. 
Below we provide a comprehensive analysis of the factors determining the maximum achievable CAR in microresonators.

In the ideal theoretical framework of SFWM in microresonators, PGR is expected to scale as $Q^3$. 
At a given pump power $P$, the photon pair flux is $\alpha \propto Q^3 P^2$, indicating $n_\mathrm{cc} \propto Q^3 P^2$. 
Simultaneously, the emission rate of noise photons (Raman and PL), denoted as $\beta$, scales as $\beta\propto QP$. 
Consequently, the accidental coincidence rate is given by $n_\mathrm{acc} \propto (Q^3P^2 + QP)^2 \Delta t$, where $\Delta t$ is the FWHM of the two-photon correlation histogram. 
Given that $\Delta t$ is proportional to the cavity lifetime (i.e., $\Delta t \propto Q$, shown in Fig. \ref{Fig:S6}b), in the low $P$ regime, we find $n_\mathrm{acc} \propto (QP)^2 \cdot Q = Q^3 P^2$. 
Under these ideal conditions, both $n_\mathrm{cc}$ and $n_\mathrm{acc}$ exhibit the same $Q$-scaling behavior. 
Therefore, the theoretical maximum CAR should, in principle, remain constant regardless of $Q$.

However, as shown in Fig. \ref{Fig:S6}c, our experimental observations reveal a discernible degradation in the maximum achievable CAR as $Q$ increases. 
We attribute this to the increased sensitivity of the energy conservation condition in the high-$Q$ regime. 
As the $Q$ factor rises, the cavity resonance linewidth narrows significantly, imposing stricter requirements on the balance between local anomalous GVD and the nonlinear frequency shifts induced by SPM and XPM. 
While the latter is power-dependent, the former (GVD) is highly sensitive to variations in fabrication process. 
This mismatch causes the actual scaling of PGR to fall short of ideal $Q^3$, i.e. effectively $\mathrm{PGR} \propto Q^{3-\epsilon}$ (where $\epsilon > 0$). 
In contrast, the generation of the accidental background---which involves the random overlap of noise photons---is not constrained by the resonance-alignment sensitivities and continues to scale as $Q^3$. 
In summary, the actual CAR scales as $1/Q^\epsilon$, explaining the observed reduction in CAR for the higher-$Q$ TM$_{10}$ mode. 
This result highlights the fundamental trade-off between cavity enhancement and the maximum achievable CAR.

\subsection{Loss budget}

System efficiency is paramount for practical quantum information applications. 
The derivation of PGR allows for accurate extraction of the overall system efficiency $\eta_\mathrm{s,i}$ for both signal and idler photons. 
For the $\mathrm{TE}_{20}$ mode (presented in the main text), the extracted efficiencies are $\eta_\mathrm{s} = 4.8\%$ and $\eta_\mathrm{i} = 6.4\%$; 
for the $\mathrm{TM}_{10}$ mode (presented in the Supplementary Materials), the extracted efficiencies are $\eta_\mathrm{s} = 5.0\%$ and $\eta_\mathrm{i} = 5.5\%$. 
This overall efficiency accounts for several discrete loss mechanisms, including photon extraction from the microresonator, fiber-to-chip edge coupling, spectral filtering, and the efficiency of the single-photon detectors. 
The efficiency of each individual component is independently measured or estimated, as detailed in Table \ref{tab:S1}. 
Notably, the overall efficiencies calculated by multiplying these individual component values are highly consistent with the results extracted directly from the PGR measurements. 

The loss budget in Table \ref{tab:S1} provides a pathway for further optimization and enables a performance estimation for deployment on satellite-based platforms. 
By utilizing a highly over-coupled microresonator, the photon extraction efficiency can be enhanced to over 80\%. 
Such an increase in $\kappa_\mathrm{ex}$ simultaneously raises the parametric threshold $P_\mathrm{th}$ and the CAR, thereby enabling both a higher maximum photon-pair flux $ \alpha^\mathrm{max}$ and improved quantum state fidelity.
Furthermore, as detailed in Note \ref{sec:package}, the fiber-chip coupling efficiency can be improved to 75\% through advanced packaging techniques. 
Consequently, the total efficiency for a single photon---from the generation within the microresonator to coupling into the fiber---can reach 60\% (2.2 dB). 
For the two-photon channel, this corresponds to a combined efficiency of 36\% (4.4 dB), which is on par with the typical transmitter efficiencies (25\%) reported in Ref. \cite{Lu:2022}.

\begin{table}[h!]
\centering
\caption{\textbf{Loss budget for PGR measurement setup.}}
\label{tab:S1}
\vspace{0.2cm}
\begin{tabular*}{0.6\textwidth}{@{}@{\extracolsep{\fill}}lcc@{~~~~}cc}
\hline\hline
\rule{0pt}{12pt}
\multirow{2}{*}{Loss source} & \multicolumn{2}{c}{TE$_{20}$} & \multicolumn{2}{c}{TM$_{10}$} \\
 & Signal & Idler & Signal & Idler \\[2pt]
\hline\rule{0pt}{10pt}%
Photon extraction   & 0.32 & 0.33 & 0.32 & 0.33 \\
Edge coupling  & 0.48 & 0.48 & 0.52 & 0.52 \\
Filtering        & 0.50 & 0.59 & 0.53 & 0.59 \\
Detection        & 0.60 & 0.60 & 0.60 & 0.60 \\[2pt]
\hline\rule{0pt}{10pt}%
Overall   & 0.046 & 0.056 & 0.053 & 0.061 \\
Overall (dB)   & 13.4 & 12.5 & 12.8 & 12.1 \\
\hline\hline
\end{tabular*}
\end{table}

\clearpage

\section{Discussion and analysis on packaging}
\vspace{0.2cm}
\label{sec:package}

\begin{figure*}[b!]
\centering
\includegraphics[width=\textwidth]{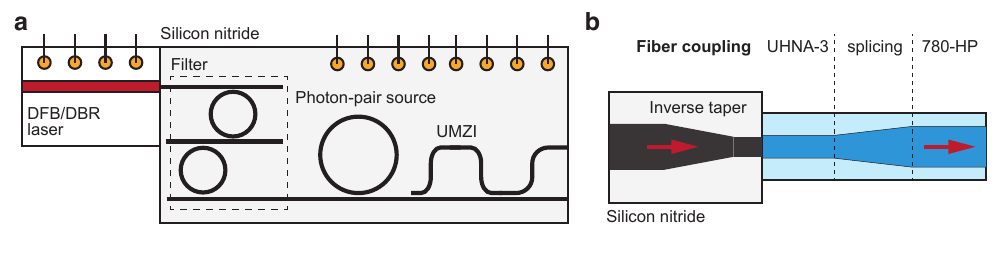}
\caption{
\textbf{Schematic of system packaging and fiber-chip coupling. }
\textbf{a}, 
Hybrid integration architecture.
A DFB laser chip is edge-coupled to the Si$_3$N$_4$ PIC. 
A series of microresonator filters suppresses sideband noise and spontaneous emission, and then the laser is self-injection-locked to the main high-$Q$ microresonator where photon pairs are generated.  
A UMZI is utilized to spatially demultiplex the signal and idler photons.
\textbf{b}, 
Fiber-chip interface. 
The generated photons are edge-coupled into a UHNA-3 fiber through optimized inverse tapers. 
The UHNA-3 fiber is then fusion-spliced to a 780-HP single-mode fiber. 
}
\label{Fig:S10}
\end{figure*}

While the near-visible photon-pair sources in this work are characterized using a free-space optical setup, the platform is inherently designed for compact, system-level packaging. 
As illustrated in the schematic in Fig. \ref{Fig:S10}a, a DFB laser chip is edge-coupled to the Si$_3$N$_4$ chip.  
Before pumping the photon-pair source, the laser output first passes through a series of microresonator filters to suppress sideband noise and spontaneous emission. 
Simultaneously, the DFB laser frequency is stabilized to the first filter microresonator via laser self-injection locking \cite{Long:25}. 
The resulting signal and idler photons are spatially separated by a UMZI based on their distinct wavelengths. 
Owing to the low thermo-optic coefficient of Si$_3$N$_4$, the chip exhibits inherent thermal stability. 
The emission frequencies of the generated photons can be stabilized to 10 MHz level using a standard thermoelectric cooler (TEC), achieving high-precision control without complex active optical feedback.
Finally, the photon pairs are edge-coupled to single-mode fibers via optimized inverse tapers \cite{ChenD:26}.

Efficient coupling to the Si$_3$N$_4$ waveguide tapers is facilitated by ultra-high-numerical-aperture (UHNA) fibers. 
Here we use UHNA-3 fibers. 
At 780 nm, the fiber's normalized frequency ($V$-number) is calculated as
$$
V = \frac{2\pi r_\mathrm{core}\mathrm{NA}}{\lambda} = 2.54,
$$
where $r_\mathrm{core} = 0.9$ $\mu$m is the fiber core radius, 
NA = 0.35 is the numerical aperture, 
and $\lambda=780$ nm is the operating wavelength. 
Although $V$ slightly exceeds the multimode cutoff ($V < 2.405$), single-mode operation is maintained by using sufficiently short, straight fiber segments. 
The mode-field radius $w$ is estimated using Marcuse’s Equation \cite{Marcuse:77}
$$
 \frac{w}{r_\mathrm{core}}\approx 0.65 + \frac{1.619}{V^{3/2}} + \frac{2.879}{V^6} = 1.06,
$$
yielding a mode-field diameter (MFD) of $2w=1.9$ $\mu$m. 
To evaluate the coupling between the UHNA-3 fiber and the Si$_3$N$_4$ inverse taper (width 300 nm, thickness 120 nm), the theoretical coupling efficiency $\eta$ is determined via the power overlap integral between the calculated TE$_{00}$ mode of the inverse taper and the Gaussian fiber mode
$$
\eta = \frac{\left|\int_A E_x(x, y) G^*(x, y) \mathrm{d}A\right|^2}{\int_A |E_x(x, y)|^2\mathrm{d}A \int_A |G(x, y)|^2\mathrm{d}A},
$$
where $E_x$ represents the dominant field component of the TE$_{00}$ mode, and $G(x, y)$ is the normalized Gaussian distribution representing the circular mode profile of the UHNA fiber with an MFD of 1.9 $\mu$m. 
This calculation yields a theoretical coupling efficiency of approximately 87\%.
To interface with standard infrastructure, the UHNA-3 fiber is fusion-spliced to a 780-HP single-mode fiber. 
By leveraging thermal core expansion during the splicing process---illustrated in Fig. \ref{Fig:S10}b---we achieve splicing efficiencies exceeding 95\%. 
Experimentally, we have measured a total fiber coupling efficiency of over 75\% per chip facet.

\clearpage

\section*{Supplementary References}
\bibliographystyle{apsrev4-1}
\bibliography{bibliography, bib_add}